\documentclass[preprint]{aastex}
\newcommand       \be           {\begin{equation}}
\newcommand       \ee           {\end{equation}}
\newcommand       \Angstrom     {\,{\rm \AA}}          
\newcommand       \eV           {\,{\rm eV}\,}
\newcommand       \K            {\,{\rm K}}
\newcommand       \cm           {\,{\rm cm}}
\newcommand       \s            {\,{\rm s}}
\newcommand       \erg          {\,{\rm erg}}
\newcommand       \kms		{\,{\rm km \, s}^{-1}}

\newcommand	  \yr		{\,{\rm yr}}
\newcommand       \nH           {n_{\rm H}}
\newcommand       \Tc           {T_{\rm c}}
\newcommand       \urad         {u_{\rm rad}}
\newcommand       \uHab         {u_{\rm Hab}^{\rm uv}}

\newcommand       \Qabs	        {Q_{\rm abs}}
\newcommand       \Qpr          {Q_{\rm pr}}

\newcommand       \Fpe	        {F_{\rm pe}}

\newcommand    	  \Frad 	{F_{\rm rad}}

\newcommand       \gtsim        {\gtrsim}

\newcommand    \Fdc        {F_{\rm drag,Coul}}
\newcommand    \Fpd        {F_{\rm pd}}

\newcommand    \sH         {s_{\rm H}}
\newcommand    \cc         {s_{\rm c}}
\newcommand    \Rarr       {R_{\rm arr}}
\newcommand    \Rpd        {R_{\rm pd}}
\newcommand    \Ro         {R_{\rm pd}^0}
\newcommand    \Rao        {R_{\rm arr}^0}
\newcommand    \tp         {\theta^{\prime}}
\newcommand    \pH         {p_{\rm H}}
\newcommand    \pHt        {p_{{\rm H}_2}}
\newcommand    \tr         {\theta_{\rm rad}}
\newcommand    \tdp        {\theta^{\prime \prime}}
\newcommand    \Fdh        {F_{\rm drag,H}}
\newcommand    \Fdhe       {F_{\rm drag,He}}
\newcommand    \tu         {\theta_{\rm u}}

\newcommand    \pet 	   {{\rm pet}}
\newcommand    \pdt 	   {{\rm pdt}}
\newcommand    \bomega     {\mbox{\boldmath$\hat{\omega}$\unboldmath}}
\newcommand    \bS	   {{\bf \hat{S}}}
\newcommand    \bx	   {{\bf \hat{x}}}
\newcommand    \by	   {{\bf \hat{y}}}
\newcommand    \bz	   {{\bf \hat{z}}}

\shorttitle{Forces on Grains}
\shortauthors{Weingartner \& Draine}

\begin{document}

\title{Forces on Dust Grains Exposed to Anisotropic Interstellar
Radiation Fields}

\author{Joseph C. Weingartner}
\affil{Physics Dept., Jadwin Hall, Princeton University,
        Princeton, NJ 08544, USA; CITA, 60 St. George Street, University of
Toronto, Toronto, ON M5S 3H8, Canada}
\email{weingart@cita.utoronto.ca}

\and

\author{B.T. Draine}
\affil{Princeton University Observatory, Peyton Hall,
        Princeton, NJ 08544, USA}
\email{draine@astro.princeton.edu}

\begin{abstract}

Grains exposed to anisotropic radiation fields are subjected to forces 
due to the asymmetric photon-stimulated ejection of particles.  These forces
act in addition to the ``radiation pressure'' due to absorption and scattering.
Here we model the forces due to photoelectron emission 
and the photodesorption of adatoms.  The ``photoelectric'' force depends on 
the ambient conditions relevant to grain charging.  We find that it is 
comparable to the radiation pressure when the grain potential is relatively 
low and the radiation spectrum is relatively hard.  The calculation of the
``photodesorption'' force is highly uncertain, since the surface physics 
and chemsitry of grain materials are poorly understood at present.  
For our simple yet plausible model, the photodesorption
force dominates the radiation pressure for
grains larger than $\sim 0.1 \micron$ exposed to starlight from OB stars.
We find that the anisotropy of the interstellar radiation field is
$\sim 10\%$ in the visible and ultraviolet.  We estimate size-dependent
drift speeds for grains in the cold and warm neutral media and find that
micron-sized grains could potentially be moved across a diffuse cloud
during its lifetime.  

\end{abstract}


\section{Introduction}

Anisotropic radiation fields are common in the interstellar medium (ISM).
Of course, the radiation is highly anisotropic at points located close to a
star or cluster or at the edge of a molecular cloud.  These include the
astronomically important photodissociation regions (PDRs), the surface layers 
of molecular clouds, where incident UV dissociates molecules.
The visible/UV radiation field in the diffuse ISM is due primarily to 
starlight; since stars are unevenly distributed, it too is anisotropic.
Dust grains in this
radiation field are subjected to torques, if their shapes are somewhat 
asymmetric, and this may be resposible for the alignment of interstellar 
grains (Draine \& Weingartner 1996, 1997), 
which is inferred from the observed polarization of starlight 
(Hiltner 1949a, b; Hall 1949; Hall \& Mikesell 1949).

Grains exposed to anisotropic radiation are subjected to forces as well 
as to torques, again with potentially important consequences.  If the 
``radiative'' forces are large enough to result in substantial gas-grain
drift, then this could lead to spatial variation in the dust-to-gas ratio.
If the drift depends on grain size, then grain coagulation could result.
Here we investigate two
potentially significant forces, due to the recoil associated with electrons
and adsorbed H atoms, which are ejected from the grain with some non-zero
probability following photon absorption.

Lafon (1990) modeled the net force due to anisotropic photoelectron emission
for grains exposed to unidirectional radiation fields and found that, for
some interstellar conditions, it substantially exceeds the radiation 
pressure, due to absorption and scattering.  Lafon considered spherical grains
and assumed that electrons are emitted only from the illuminated hemisphere.
However, interstellar grain radii can be smaller than the photon attenuation
length, so that electrons are also emitted from the non-illuminated 
hemisphere.  In \S \ref{sec:fpe}, 
we revisit the calculation of photoelectric recoil 
forces on dust grains, employing the prescription of Kerker \& Wang (1982)
for estimating the anisotropy in photoelectron momentum.

In \S \ref{sec:fpd}, 
we consider the force due to the anisotropic photodesorption of 
adsorbed H atoms.  An adatom starts out in the gas phase, collides with 
and sticks to the grain surface, and can be removed when an absorbed photon 
breaks its bond to the surface.  Of course, there are many other grain
surface processes, besides adsorption and photodesorption, and these too
could play important roles in determining the emission anisotropy.  
Since interstellar grains have not yet been well characterized, the details
of their surface physics are unknown.  We adopt a simple, yet plausible,
model and evaluate the ``photodesorption'' force for this one case.

Following the investigations of the photoelectric and photodesorption forces,
we discuss, in \S \ref{sec:drag}, how surface processes can affect the drag
force experienced by drifting grains.  In \S \ref{sec:anisotropy}, we 
estimate the anisotropy of the interstellar radiation field.  
In \S \ref{sec:drift_diffuse}, we 
estimate drift speeds for grains in the diffuse ISM, tying together the 
results from the previous sections.  In a separate paper (Weingartner \& 
Draine 2001c), we will investigate gas-grain drift in PDRs.

\section{Radiation Pressure Force}

We  generally normalize our results for the ``photoelectric'' and
``photodesorption'' forces to the radiation pressure force, in order
to see whether or not they are significant.  The force due to radiation 
pressure is 
\be
\Frad = \pi a^2 \langle Q_{\rm pr}\rangle \Delta \urad~~~,
\ee
where $a$ is the grain radius (we assume spherical grains in all of our
calculations) and
$c\Delta\urad$ is the net energy flux in the radiation field
($\Delta\urad=\urad$, the total energy density, 
for a unidirectional radiation field
and $\Delta\urad=0$ for an isotropic radiation field).  The 
spectrum-averaged radiation pressure efficiency factor is given by
\be
\langle Q_{\rm pr}\rangle \equiv 
\frac{
        \int_0^{\nu_{\rm max}} 
        \left[ Q_{\rm abs} + 
                Q_{\rm sca}(1-\langle\cos\theta\rangle)
        \right]
        u_\nu d\nu
        } 
        {\int_0^{\nu_{\rm max}} u_\nu d\nu
        }~~~,
\ee
where $\Qabs\pi a^2$ is the absorption cross section,
$Q_{\rm sca}\pi a^2$ is the scattering cross section,
$\langle\cos\theta\rangle$ is the usual scattering asymmetry factor, and
$\nu_{\rm max}$ is the highest frequency in the radiation field.  We will
take $h\nu_{\rm max}=13.6 \eV$, because H~I regions are highly opaque to 
H-ionizing radiation.  We evaluate $Q_{\rm abs}$, $Q_{\rm sca}$, and
$\langle\cos\theta\rangle$ using a Mie theory code derived from BHMIE 
(Bohren \& Huffman 1983).  We consider carbonaceous and silicate compositions
and take optical properties from Draine \& Lee (1984), Laor \& Draine (1993),
and Li \& Draine (2001), as described by Weingartner \& Draine (2001a).

We consider two types of radiation spectrum.  To characterize regions
located near a single star,
we adopt a blackbody spectrum (cut off at $13.6 \eV$), with
color temperature $T_{\rm c}$ and dilution factor $W$, so that 
the energy density per unit frequency interval $u_\nu =
4 \pi W B_\nu (T_{\rm c}) /c$.  It is convenient to characterize the
radiation intensity by $G \equiv
\urad^{\rm uv} / \uHab$, where $\urad^{\rm uv}$ is the 
energy density in the radiation field between $6 \eV$ and $13.6 \eV$
and $\uHab = 5.33 \times 10^{-14} \erg {\cm}^{-3}$ is the 
Habing (1968) estimate of the starlight energy density in this 
range.\footnote{
	For comparison, the interstellar radiation field estimated by
	Draine (1978) has $u=8.93\times10^{-14}\erg\cm^{-3}$ between
	6 and 13.6 eV, or $G=1.68$.
	}

For the diffuse ISM, we adopt
the average interstellar radiation field (ISRF) spectrum in
the solar neighborhood, as estimated by
Mezger, Mathis, \& Panagia (1982) and  Mathis, Mezger, \& Panagia (1983);
see Weingartner \& Draine (2001b) for a convenient representation of the 
ISRF.  The total energy density in the ISRF is
$u=8.64\times10^{-13}\erg\cm^{-3}$, with
$u_{\rm rad}^{\rm uv}=6.07\times10^{-14}\erg\cm^{-3}$ in the
6-13.6 eV interval, or $G=1.13$.

\begin{figure}
\epsscale{1.00}
\plotone{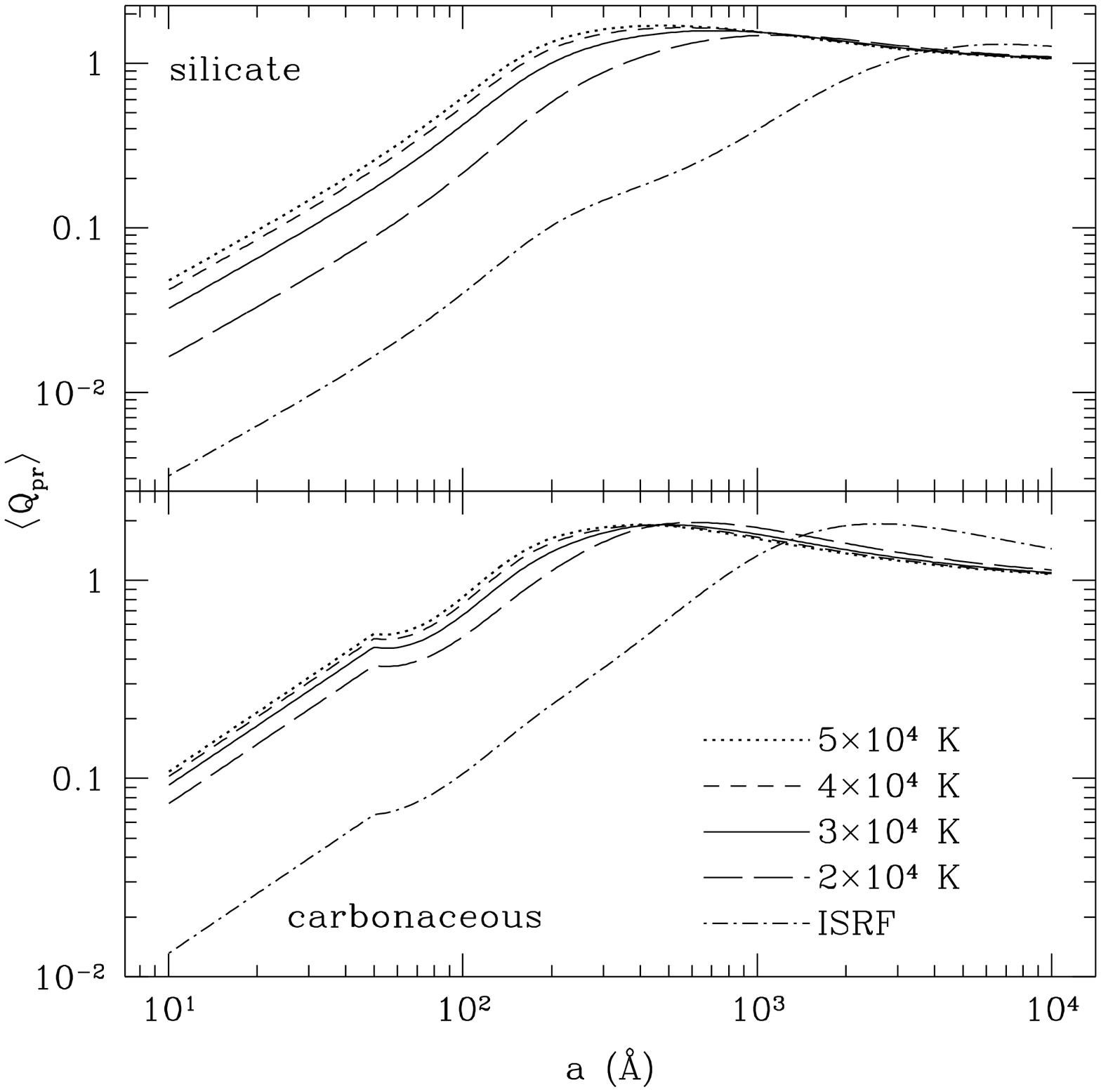}
\caption{
\label{fig:qpr}
	Radiation pressure efficiency factors for neutral carbonaceous and 
silicate grains, averaged over the interstellar radiation field (ISRF) and 
blackbody spectra with indicated color temperatures (cut off at $13.6 \eV$).
The kink at $a=50 \Angstrom$ results from the Li \& Draine (2001) prescription
for blending PAH and graphite optical properties. 
        }
\end{figure}

In Figure \ref{fig:qpr} 
we display $\langle \Qpr \rangle$, as a function of grain size $a$, for 
the ISRF and blackbody spectra with various values of $T_c$.  We consider
grains with $a$ as large as $1 \micron$, even though the classic MRN 
size distribution (Mathis, Rumpl, \& Nordsieck 1977)
only extends up to $a=0.25 \micron$.  The size distributions of 
Weingartner \& Draine (2001a) contain non-negligible amounts of mass in 
grains with $a \sim 1 \micron$, and we will find that the radiative forces
are relatively large for such grains; thus, they could be important in some
interstellar environments.  We consider grains as small as $a = 10 
\Angstrom$, equal to the value adopted by Weingartner \& Draine (2001b)
for the electron escape length, the typical distance that an excited 
electron travels within a grain before losing its energy.  This choice is
somewhat arbitrary, but this does not matter, since the radiative forces
are inconsequential for grains with $a \sim 10 \Angstrom$.   

\section{``Photoelectric'' Force \label{sec:fpe}}

\subsection{Model}

We adopt the photoemission model of Weingartner \& Draine (2001b).  
The photoemission rate and photoelectron energy both depend on the grain 
charge, so it is necessary to evaluate the charge distribution $f_Z(a)=$
the probability that the grain has charge $Ze$, where $e$ is the proton charge.
The charge distribution is set primarily by a balance between electron 
loss via photoemission and gain via accretion from the gas.  Charging by this
process is determined mainly 
by the parameter $G \sqrt{T}/n_e$ ($T$ is the gas temperature and
$n_e$ is the electron density), with an additional mild dependence on $T$
(Bakes \& Tielens 1994; Weingartner \& Draine 2001b).  

The photoelectric force on a grain is given by
\be
\label{eq:fpe}
\Fpe (a) = \sum_Z f_Z(a) \left[ F_{\rm pe,v}(a,Z) + F_{\rm pe,a}(a,Z) 
\right]~~~,
\ee
where $F_{\rm pe,v}(a,Z)$ is the force due to electrons emitted from the 
valence band.  When $Z<0$, the $-Z$ ``attached'' electrons occupy energy 
levels above the valence band, if the latter is full in the neutral;
$F_{\rm pe,a}(a,Z)$ is the force due to the emission of these attached 
electrons.  

The contribution from the photoemission of valence electrons is 
\be
F_{\rm pe,v}(a,Z) = \pi a^2 \int_{\nu_\pet}^{\nu_{\rm max}} d\nu A Y \Qabs
\frac{c u_{\nu}} {h\nu} \sqrt{2m_e} \int_{E_{\rm min}}^{E_{\rm max}} dE
f_E(E) \sqrt{E} \, S(Z,a,E)~~~,
\ee
where the photoemission threshold energy $h\nu_\pet$ is the minimum energy
that an absorbed photon must have for an electron to be ejected from the
valence band, $E_{\rm min}$
($E_{\rm max}$) is the minimum (maximum) photoelectron kinetic energy
(at infinity), and $f_E(E)$ is the photoelectron kinetic energy (at infinity)
distribution; see Weingartner \& Draine (2001b) for prescriptions for 
evaluating these quantities.
The emission asymmetry factor $A(h\nu,a)$
measures the asymmetry in the emission of photoelectrons over the
grain surface, the recoil suppression factor $S(Z, a, E)$
accounts for electron emission in directions other than the surface normal,
and $m_e$ is the electron mass.

When $Z<0$, the contribution from photodetachment is 
given by
\be
F_{\rm pe,a}(a,Z) = \int_{\nu_\pdt}^{\nu_{\rm max}} d\nu A \sigma_\pdt (\nu)
\frac{c u_{\nu}} {h\nu} \sqrt{2 m_e (h\nu - h\nu_\pdt + E_{\rm min})}
S(Z, a, E = h\nu - h\nu_\pdt + E_{\rm min})~~~,
\ee
where the photodetachment threshold energy $h \nu_\pdt$ is the minimum 
energy that an absorbed photon must have for an attached electron to be 
emitted from the grain and $\sigma_\pdt (\nu)$ is the photodetachment cross
section.

\begin{figure}
\epsscale{1.00}
\plotone{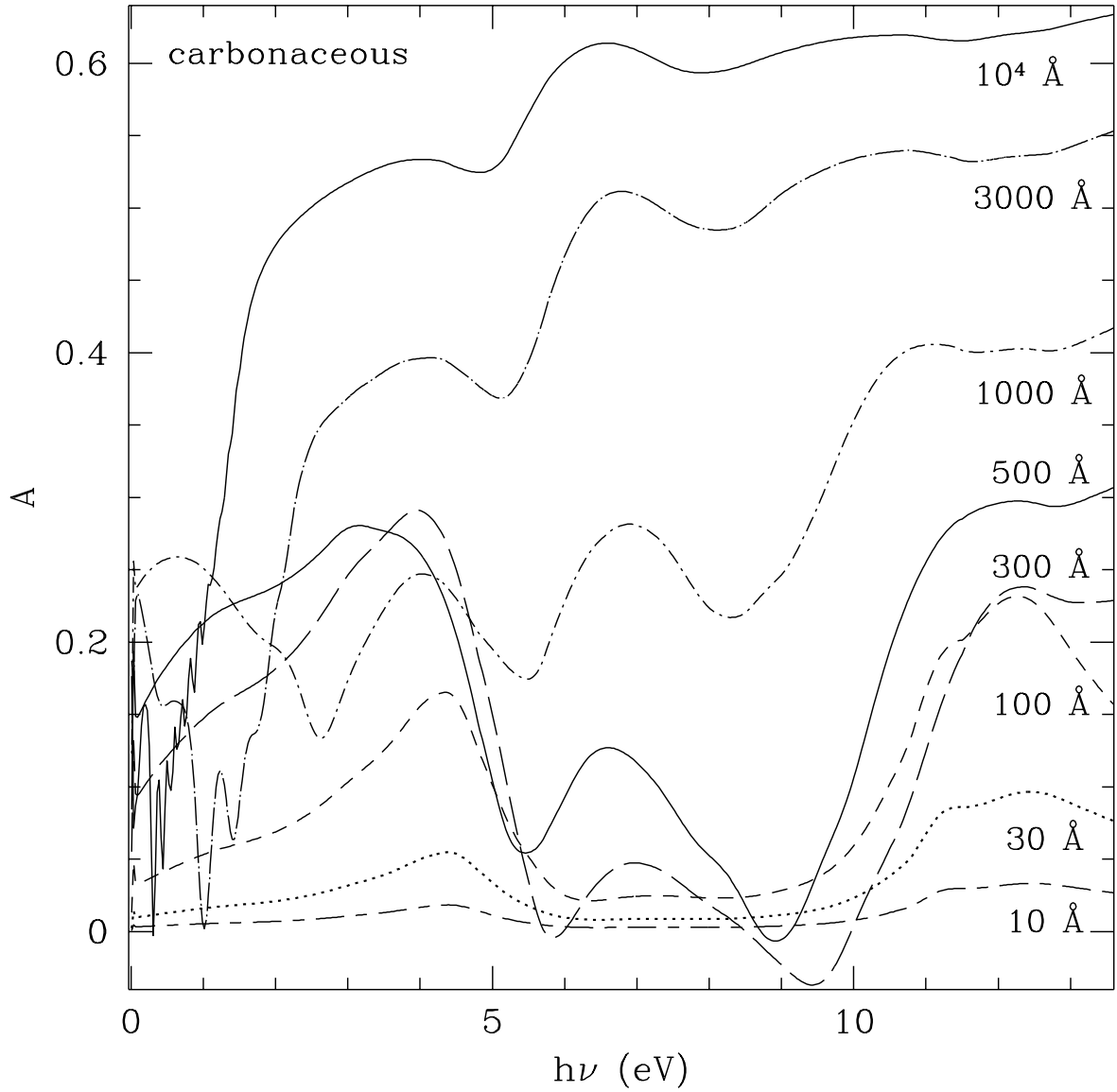}
\caption{
        The asymmetry factor $A$ as a function of the incident photon
        energy $h \nu$, for carbonaceous grains.  The grain radius is
        indicated for each curve.
        }
\label{fig:asymmetry_factor_C}
\end{figure}

\begin{figure}
\epsscale{1.00}
\plotone{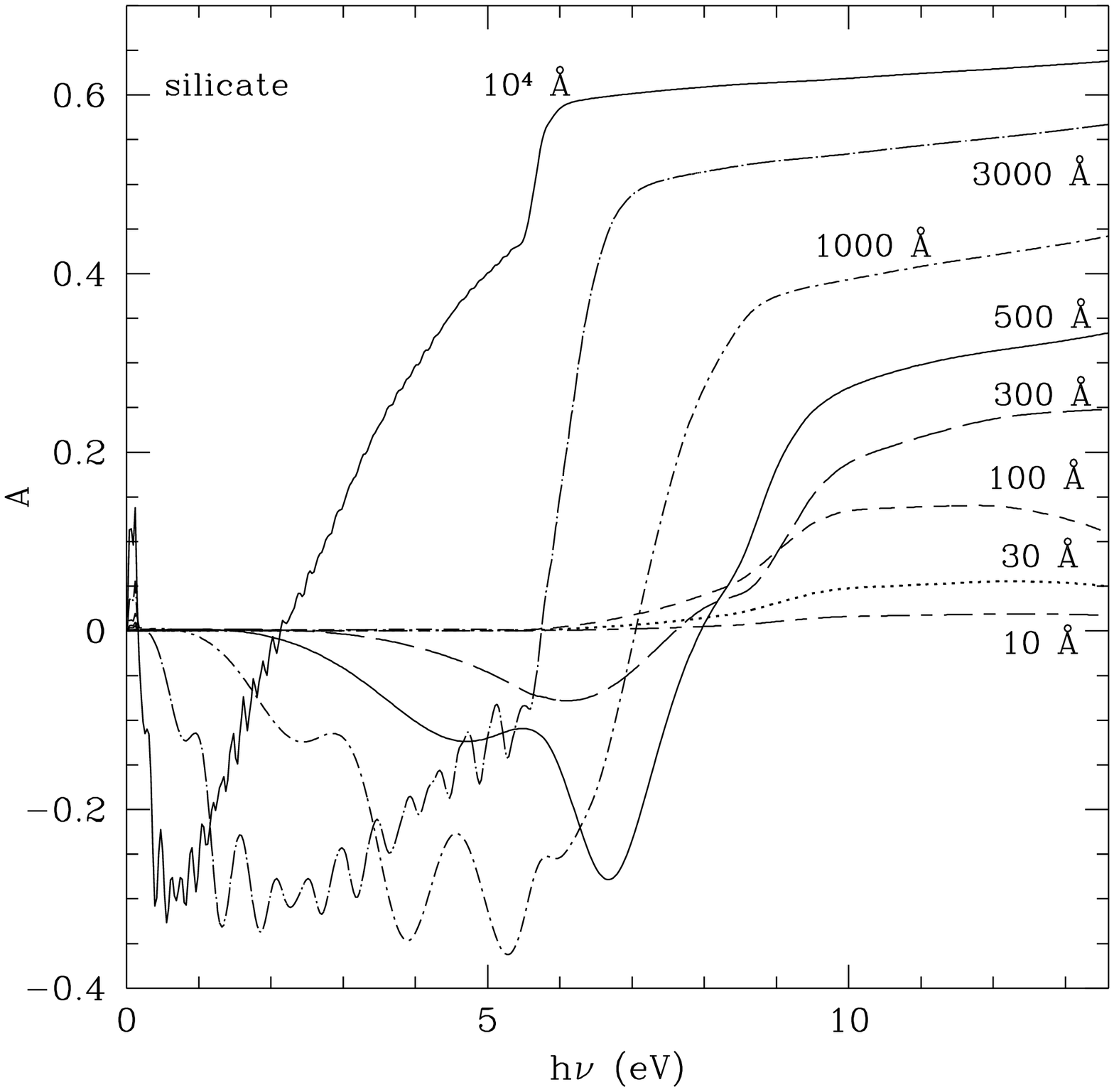}
\caption{
        Same as Figure \protect{\ref{fig:asymmetry_factor_C}},
        but for silicate grains.
        }
\label{fig:asymmetry_factor_sil}
\end{figure}

We follow the simple prescription of Kerker \& Wang (1982) for estimating
$A$.  The probability of photoemission from any site on the surface is taken
to be proportional to the electric intensity ${\left| \bf{E} \right|}^2$ 
just below the surface at that point.  Thus, 
\be
A(h\nu, a) = \frac{-\int_0^{\pi} \sin\theta \cos\theta{\left| \bf{E}(\theta)
 \right|}^2 \, d\theta}{\int_0^{\pi} \sin\theta {\left| \bf{E}(\theta)
\right|}^2 \, d\theta}~~~,
\ee
where $\theta$ is the polar angle with respect to the direction of the 
incident radiation.  The electric field $\bf{E}(\theta)$ is
evaluated using Mie theory (Bohren \& Huffman 1983).\footnote{We adopt   
dielectric functions as in Weingartner \& Draine (2001b).}  In Figures 
\ref{fig:asymmetry_factor_C} and \ref{fig:asymmetry_factor_sil} we
display the asymmetry factor $A(h\nu, a)$ for grains exposed to  
unidirectional radiation fields as a function of incident photon energy
for various grain sizes for carbonaceous and silicate composition, 
respectively.
Generally, $A(h\nu, a)$ equals the value for a unidirectional field times
$\Delta u_{\rm rad} / u_{\rm rad}$. 
Curiously, $A < 0$ for some values of ($h\nu$, $a$).

We assume that the electrons emerge symmetrically with respect to the local
surface normal, with a ``cosine-law'' angular distribution
(i.e.~the emission rate at angle $\psi$ with respect to the surface
normal $\propto \sin\psi \cos\psi \, d\psi$).  For an uncharged grain, this
would imply $S=2/3$.  When the grain is charged, the electron escapes
on a hyperbolic trajectory, so that when it is at infinity
its velocity vector makes an angle $\psi_{\infty}$ with respect to
the surface normal which differs from the corresponding angle at the 
surface, $\psi$.  Taking this into account, we find
\begin{eqnarray}
S & \equiv & \frac{\int_0^{\pi/2} d\psi \sin\psi \cos\psi \cos\psi_{\infty}}
{\int_0^{\pi/2} d\psi \sin\psi \cos\psi} \nonumber\\ 
& = & \frac{2}{3} \sqrt{1+b} -\frac{b}{2} +\frac{b^2 (1+b/2)}
{2 (1+b)} \left[ \ln \left| 1 + \frac{2}{b} \right| + \frac{1}{2} \ln
\left(\frac{1+b/2 - \sqrt{1+b}}{1+b/2 + \sqrt{1+b}}\right)\right] + \frac
{b^2}{2 \sqrt{1+b}}~~~,
\end{eqnarray}
\be
b\equiv \frac{(Z+1)e^2}{a E}~~~,
\ee
where $Ze$ is the grain charge prior to photoelectron emission.
For $-1 \le b \le 10^4$, the simpler expression
\be
S \approx \frac{1.107}{0.4669 + (b+1.421)^{0.5043}}
\ee
is accurate to within $0.6 \%$; we adopt this expression in our
computations.  When an electron tunnels across the Coulomb barrier of
a negatively charged grain, $b < -1$ and the above classical derivation fails, 
since the grain surface is a classically forbidden location.  We simply take
$S=1$ when $b < -1$.  

\subsection{Computational Results}
\label{sec:fperesults}

The photoelectric force depends on the grain charge distribution, and hence
on the parameters $G \sqrt{T}/n_e$ and $T$.  
In Figures \ref{fig:fpegra} and \ref{fig:fpesil}, $\Fpe$ is plotted for
carbonaceous and silicate grains, respectively, for $T=10^2$ and $10^3 \K$ and
three values of $G \sqrt{T}/n_e$, ranging from 
$10^3$ to $10^5 \K^{1/2} \cm^{3}$.  For these cases, we adopt a blackbody
spectrum with $\Tc = 3 \times 10^4 \K$, cut off at $13.6 \eV$.  
We also provide results for conditions appropriate for the cold neutral
medium (ISRF, $T=100\K$, $n_e=0.03 {\cm}^{-3}$, $G\sqrt{T}/n_e = 380$)
and the warm neutral medium (ISRF, $T=6000\K$, $n_e=0.03 {\cm}^{-3}$, 
$G\sqrt{T}/n_e= 2900$). 

Note that for a given value of $G\sqrt{T}/n_e$,
the results do not change much as the temperature 
increases from $10^2$ to $10^3 \K$.
Also, $\Fpe / \Frad$ decreases for higher $G \sqrt{T} /n_e$
because the higher ionization potentials quench the photoemission.  

For the truncated blackbody spectrum, there is
no systematic increase or decrease of $\Fpe / \Frad$ over the full range of
grain sizes, because the increase in the anisotropy of emitted electrons with 
grain size roughly compensates for the decrease in photoelectric yield. 
The sharp minimum near $300 \Angstrom$ in Figure \ref{fig:fpegra}
results from a local plateau in the variation of the 
anisotropy with grain size, which lies near the maximum of $F_{\rm{rad}}$.
For the ISRF,
the photoelectric force becomes less important, compared with the radiation 
pressure, as the grain size increases beyond $\sim$200\AA.
As seen in Figure \ref{fig:qpr}, for the relatively soft ISRF,
the radiation pressure increases rapidly 
as $a$ increases from $10 \Angstrom$ to $1 \micron$.
The total energy density between $0$ and $13.6 \eV$ is about an order
of magnitude greater for the ISRF than for a blackbody with $\Tc = 3 \times
10^4 \K$ and $G=1$.
For $a = 10 \Angstrom$, $\Frad$ is comparable for the two spectra,
but is about an order of magnitude larger for the ISRF when $a = 1 \micron$.
Since $G= 1.13$ for the ISRF, $\Fpe$ is comparable for the two spectra.
  
In Figure \ref{fig:fpetc} we show the 
effect of varying $\Tc$ over a range from $2 \times {10}^4$ to 
$5 \times {10}^4 \K$, characteristic of hot stars, using carbonaceous
grains and
$G\sqrt{T}/n_e = 10^3 \K^{1/2} \cm^3$, $T = 100 \K$ as an example.  

\section{``Photodesorption'' Force \label{sec:fpd}}

\subsection{Introduction and Model}

In the process of adsorption, gas-phase atoms or molecules collide with 
and stick to a solid surface, forming either a weak van der Waals bond 
(physisorption) or a stronger chemical bond (chemisorption).  Adsorbed 
atoms can be removed from interstellar grains by various desorption 
processes:  1) thermal desorption, in which a thermal fluctuation in the
grain breaks the adsorption bond, 2) photodesorption, in which an 
absorbed photon breaks the bond, and 3) exothermic chemical reaction 
between two adatoms.  If photodesorption is important, we expect
a non-zero net recoil force in an anisotropic radiation field.

The details of adsorption on grain surfaces are poorly known at
present.  In particular, it is unclear whether adsorbed hydrogen atoms
can diffuse across the surface, either by thermal barrier hopping or
quantum mechanical tunneling.  In their classic study of H$_2$
formation on grains, Hollenbach \& Salpeter (1971) assumed that H
physisorbs with binding energy $\sim 0.03 \eV$, implying high
mobility.  Physisorption occurs on ice surfaces, so that this
description is likely valid deep inside molecular clouds, where grains
are covered with ice mantles.  However, grain temperatures in PDRs are
high enough that any ice mantles have sublimed, and ice features  
(e.g. at $3.1 \micron$ for H$_2$O ice) are not observed in the 
extinction curves for the diffuse ISM.  Tielens \&
Allamandola (1987) discuss experimental evidence that H chemisorbs on
graphite and amorphous silicate surfaces, with binding energy in the
1--2$\eV$ range.  Also, theoretical studies (Mendoza \& Ruette
1989; Fromherz, Mendoza, \& Ruette 1993) of H adsorption on graphite
find chemisorption with binding energies in this range.  For such
large binding energies, thermal diffusion is negligible.  The tunneling
rate is approximately 
given by the oscillation frequency of the atom in the chemisorption
potential times the barrier penetration factor, so that
\be
\tau_t \sim 10^{-13} \, {\rm s} \, \exp \left[ \frac{2 d}{\hbar} \left(
2 m E_b \right)^{1/2} \right]~~~,
\ee
where $\tau_t$ is the tunneling timescale,
$d$ is the distance between binding sites, $m$ is the mass of the 
adatom, and $E_b$ is the binding energy.  The argument of the 
exponential is a large number; if ($E_b$, $d$) = ($1 \eV$, $1 \Angstrom$),
then $\tau_t \sim 13 \,$days for H adatoms.  
If $d$ is increased to $2 \Angstrom$, then
$\tau_t \sim 4 \times 10^{17} \,$years.  
We assume that no diffusion occurs, primarily because this case is easier
to model.  
Once a binding site is occupied by a chemisorbed H atom, another H can
physisorb on top of the first (Tielens \& Allamandola 1987).  An atom in
such a  physisorption state can easily migrate to neighboring sites;
also, if it can overcome an activation energy, it can react with the
underlying H atom.  

In our model, we assume that each surface site is a chemisorption
site, from which adatoms cannot diffuse.  
For simplicity, we also ignore gas-phase species other than H.
When an H atom from the gas arrives at an empty site, it sticks with
probability $\sH$; otherwise it is assumed to undergo perfectly elastic
specular reflection.  Since
the binding energy in the chemisorption site greatly exceeds the
kinetic energies of the atoms in the gas, we take $\sH = 1$.  Also,
we make the simplifying assumption that an H atom arriving at an
occupied site either combines with the resident chemisorbed atom, with
probability $\cc$, or else undergoes elastic specular reflection.  This
approximation is most reasonable for high gas temperatures, when the
kinetic energies of the gas-phase atoms exceed the physisorption
energy and reaction activation barrier.  
We also assume that H$_2$ molecules leave the grain immediately upon
formation.  

To calculate the net force on a grain we must evaluate, as a function
of location on the grain surface, the average 
rate at which momentum is transferred to the grain per surface site.  
Momentum transfer occurs when an atom is photodesorbed, a newly-formed 
molecule leaves the grain, a gas-phase atom reflects off of the grain, or
a gas-phase atom sticks to the grain (either at an unoccupied site, where
it remains until it is removed by photodesorption or molecule formation, or
at an occupied site, where it immediately forms a molecule, which leaves
the grain).  

We denote the rate at which gas phase H atoms arrive at a
site by $\Rarr$.  The rate at which incident H sticks per
empty site is given by $\sH \Rarr$ and the rate at which H$_2$
is formed per occupied site is given by $\cc \Rarr$.  
When a grain drifts with respect to the gas the arrival rate $\Rarr$
varies over the grain surface; of course, this results in the drag force.
Baines, Williams, \& Asebiomo (1965) calculated the drag, assuming that all
arriving atoms undergo specular reflection.  The surface processes
considered here substantially complicate the drag calcuation and the
variable $\Rarr$ complicates the net force due to photodesorptions.  
We will call the net force on a non-drifting grain, due to all of the
processes mentioned above, the ``photodesorption'' force.  In \S 
\ref{sec:drag} we will discuss how the drag force is modified when the 
surface processes are taken into account.  

\subsection{Photodesorption Rate \label{sec:pdrate}}

The rate at which H is photodesorbed per occupied site is denoted by $\Rpd = 
\Ro \Psi (\tp)$, where $\Ro$ is the photodesorption rate averaged over the 
grain surface ($\approx$ the incident photon flux density times the 
photodesorption cross section) and $\Psi (\tp)$ describes the variation in
photodesorption rate as a function of the angle $\tp$ with respect to the 
direction $\bS$ of net energy flux.  

The photodesorption cross sections are not
known.  There is a substantial literature on experimentally determined
cross sections for simple molecules on a variety of substrates, which
is summarized in the review by Franchy (1998).  For example, for NO
and photon energies $\sim 6 \eV$ or less, the cross sections range
from $\sim 10^{-22} \cm^2$ to $2 \times 10^{-17} \cm^2$.  The cross
section increases with progression through the following sequence of
substrates:  platinum metals, noble metals, semiconductors, oxides.  
The differences are due to variation in the efficiency with which the
substrates quench the excited states.  

Hellsing et al.~(1997) studied the photodesorption of K from graphite
for photon energies $\le 5 \eV$ and found a maximum cross section
of $\approx 2 \times 10^{-20} \cm^2$.  However, for these photon
energies the desorption is substrate-induced.  That is, a bulk
excitation (e.g. a photoelectron) travels to the adsorbed atom and
excites it.  

Photons with adequately high energy can directly excite the adsorbed atoms.  
We estimate the photodesorption cross sections due to this direct process,
for H on graphite and silicates, by considering the photodissociation of 
the CH and OH molecules.  Photodissociation cross sections for CH and OH
have been calculated by van Dishoeck (1987) and van Dishoeck \& Dalgarno 
(1984), respectively.  The cross sections reach values as high as a few 
$\times 10^{-17} \cm^2$, and are appreciable for photon energies 
$\ge 6 \eV$.  Roberge et al. (1991) found 
photodissociation rates of $8.6 \times 10^{-10} 
{\rm s}^{-1}$ and $3.5 \times 10^{-10} {\rm s}^{-1}$ for CH and OH, 
respectively, for the Draine (1978) radiation field.  
In the absence of more relevant experimental evidence, we simply set
the photodesorption rates equal to the photodissociation rates.  Also, in
our crude approximation, we ignore variations in the shape of the 
radiation spectrum, and simply take the photodesorption rate proportional 
to $ u_{\rm rad}^{\rm uv}$,
the total energy density between 6 and $13.6 \eV$.  
Since $G = 1.68$ for the Draine radiation
field, we take $\Ro = 5 \times 10^{-10} G \s^{-1}$ for carbonaceous grains
and $\Ro = 2 \times 10^{-10} G \s^{-1}$ for silicates.  

\begin{figure}
\epsscale{1.00}
\plotone{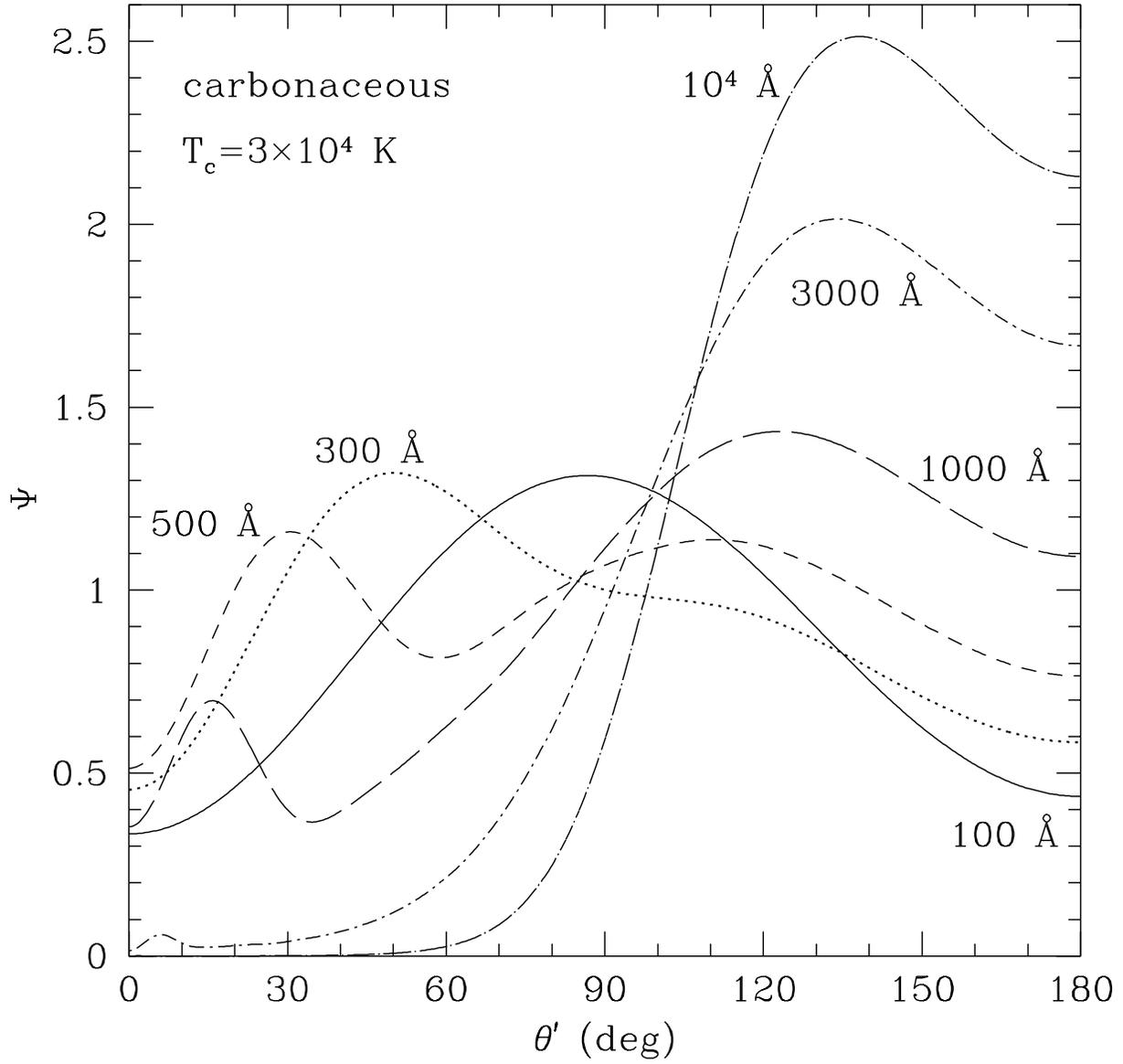}
\caption{
\label{fig:psi_gra}
Normalized rate for photodesorption 
$\Psi(\tp)$ for carbonaceous grains, a blackbody radiation spectrum (cut off at
$13.6 \eV$) with $\Tc = 3 \times 10^4 \K$, and several values of grain size 
$a$, as labeled.
        }
\end{figure}

\begin{figure}
\epsscale{1.00}
\plotone{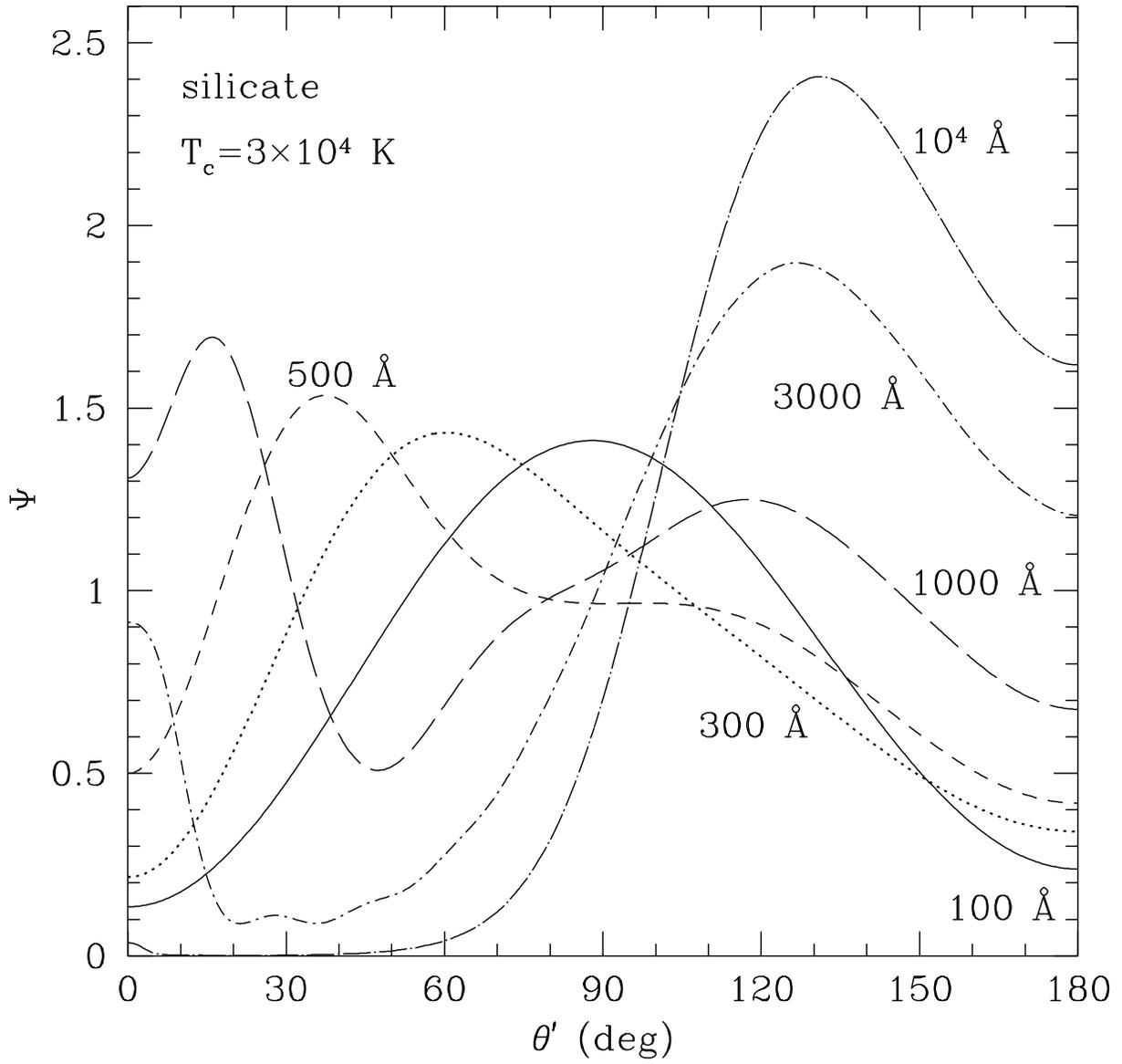}
\caption{
\label{fig:psi_sil}
Same as fig. \ref{fig:psi_gra}, but for silicate.
        }
\end{figure}

To evaluate $\Psi(\tp)$ we assume that, for any point on the grain surface,
$\Rpd$ is proportional to the electric intensity $|{\bf E_e}|^2$
just above the surface at that point.  Thus,
\be
\label{eq:psidef}
\Psi (\tp) = \frac{2 \int_{6 \eV /h}^{13.6 \eV /h} d\nu \, u_{\nu} 
{\left| \bf{E_e(\tp)} \right|}^2}{\int_0^{\pi} d\xi \sin\xi
\int_{6 \eV /h}^{13.6 \eV /h} 
d\nu \, u_{\nu} {\left| \bf{E_e(\xi)} \right|}^2}~~~.
\ee
In Figures \ref{fig:psi_gra} and
\ref{fig:psi_sil}, we display $\Psi(\tp)$ for carbonaceous and silicate
grains, respectively, for a unidirectional radiation field and  
several values of grain radius $a$.  Here we assume a blackbody spectrum
(cut off at $13.6 \eV$) with $\Tc = 3 \times 10^4 \K$.  For $a < 100
\Angstrom$, $\Psi(\tp)$ looks like the $100 \Angstrom$ curve, with greater
symmetry for smaller grains.  Generally,
\be
\Psi(\tp) = \frac{(u_{\rm rad} - \Delta u_{\rm rad}) + \Delta u_{\rm rad}
\Psi^0(\tp)}{u_{\rm rad}}~~~,
\ee
where $\Psi^0(\tp)$ is the value for a unidirectional field.

\subsection{Surface Site Occupation Fraction}

The momentum transfer rates depend on the occupation fraction $f$, i.e.~the
fraction of the time that a site is occupied by an H atom.  If the grain's
surface is fixed with respect to the radiation, then $f$ is a function of
the angle $\tp$ with respect to $\bS$.  However, the polarization
of starlight implies that grains spin, with their spin axes statistically
aligned (presumably with the magnetic field).  Purcell (1979) noted that 
systematic torques likely act on grains, resulting in  rotational speeds
orders of magnitude greater than what would be expected from equipartition
arguments.  The spin period is much shorter than the timescales
on which the surface processes occur, so that $f$ is a function of the angle
$\theta$ with respect to the spin axis $\bomega$.

We evaluate $f(\theta)$ by balancing the the rate at which H is removed 
from an occupied site by photodesorption and H$_2$ formation against the 
rate at which H is added to an empty site when arriving gas-phase atoms stick:
\be
f \left( \bar{R}_{\rm pd} + \cc \bar{R}_{\rm arr} \right) = \left( 1 - f
\right) \sH \bar{R}_{\rm arr}~~~,
\ee
where bars denote averaging over the spin.  This yields
\be
f(\theta) = \sH \left[ \sH + \cc + 
\alpha \bar{\Psi}(\theta) \right]^{-1}~~~.
\ee
The parameter $\alpha$ is defined by
\be
\alpha \equiv \frac{\Ro}{\Rao}~~~,
\ee
where the H arrival rate for a non-drifting grain is given by 
\be
\Rao = \frac{1}{2} \nH l^2 (\pi \beta)^{-1/2}~~~;
\ee 
$\nH$ is the H number density, $l^2$ is the area 
of a surface site, and $\beta = m_p / 2 k T$.  
Of course, 
\be
\bar{\Psi}(\theta) = \frac{1}{2 \pi} \int_0^{2 \pi} d\phi \Psi(\tp)~~~,
\ee
where $\phi$ is the azimuthal angle about $\bomega$.

Lazarian \& Draine (1999a,b) have recently suggested that grains sometimes
undergo rapid ``flipping'' 
of the principal axis of largest moment of inertia 
(fixed in body coordinates), alternating between
being parallel or antiparallel to the angular momentum vector.
We consider the limit where flips occur instantaneously, so that
\be
f(\theta) = \sH \left\{ \sH + \cc + \frac{\alpha}{2}
\left[\bar{\Psi}(\theta) + \bar{\Psi}(\pi - \theta) \right] \right\}^{-1}~~~.
\ee
Although rapid flipping does not seem likely for suprathermally rotating 
grains with $a \gtsim 0.1 \micron$ (because of the large energy barrier that
must be overcome in order to flip), the calculation of the photodesorption 
force in this case is instructive.  Thus,
we will consider both flipping and non-flipping grains.

\subsection{Evaluation of the Photodesorption Force}

We assume that the direction of net energy flux $\bS$ lies at
angle $\tr$ with respect to the grain's spin axis 
$\bomega$\footnote{We consider $\tr$ to be constant; i.e. we 
ignore the likely precession of $\bomega$ about the magnetic 
field.}.  
We adopt a coordinate system with the z-axis lying along 
$\bomega$ and the x-axis oriented so that $\bS$ lies
in the x-z plane.  The z-axis serves as the polar axis for spherical 
coordinates, with the x-axis the reference for azimuth; thus the spherical
coordinates for $\bS$ are $(\tr, 0)$.  The ``radiation'' angle
$\tp$ is given by
\be
\cos \tp = \sin \tr \sin \theta \cos \phi + \cos \tr \cos \theta~~~.
\ee

We assume that both the photodesorbed H atoms and newly-formed H$_2$ 
molecules emerge symmetrically with respect to the
local surface normal, with a ``cosine-law'' angular distribution, as 
adopted for photoelectrons.  The momenta of the
outgoing atoms and molecules are denoted by $\pH$ and $\pHt$
respectively.  We ignore polarization of the outgoing atoms and molecules 
when the grain has non-zero charge.  The momentum transfer, per 
surface site, due to each of
the contributing processes is directed radially inwards, with rates
as summarized in Table \ref{tab:momtransfer}.  The factors of $2/3$ appearing
in Table \ref{tab:momtransfer} are due to the cosine-law angular 
distribution.  For reflections and sticking, the term in brackets is the 
probability that the process will occur; we include the absorption of 
the gas-phase momentum of an arriving atom which will immediately undergo 
H$_2$ formation in the ``sticking'' category.  The probability that the 
component of a gas H atom's velocity along the surface normal lies between
$v$ and $v + dv$ is given by $P(v) dv = (\beta / \pi)^{1/2} \exp(-\beta v^2)
dv$.

Integrating the momentum transfer rates over all surface sites yields the
photodesorption force:
\be
\frac{\bf \Fpd}{\gamma \pi a^2} = - \bz \int_0^{\pi} d\theta
\sin\theta \cos\theta f(\theta) \left[ C + \bar{\Psi}(\theta) \right]
- \bx \int_0^{\pi} d\theta \sin^2 \theta f(\theta) 
\frac{1}{2 \pi} \int_0^{2 \pi} d\phi \cos\phi \Psi(\tp)~~~,
\ee
where 
\be
\gamma \equiv \frac{4 \Ro \pH}{3 l^2}
\ee
and
\be
C \equiv \frac{1}{\alpha} \left[ \cc \left(\frac{2 E_{{\rm H}_2}}{E_{\rm H}}
\right)^{1/2} + \frac{3}{4} \left( \frac{\pi k T}{E_{\rm H}} \right)^{1/2}
\left( \sH -\cc \right) \right]~~~.
\ee

Since the H arrival rate does not depend on surface location for a 
non-drifting grain, the contributions due to H$_2$ formation, reflections,
and sticking depend only on $f(\theta)$ and are thus directed along 
$\bz = \bomega$; 
they are all accounted for in the constant $C$.  For rapidly flipping 
grains, $f(\theta) = f(\pi - \theta)$, so that only photodesorption 
contributes to the momentum transfer; this is also true for
non-flipping grains when $\tr = 90^{\circ}$.  In these cases, 
for given 
\{$\gamma$, $\sH$, $\cc$, radiation field\}, 
the maximum force occurs when 
$\nH \rightarrow \infty$, so that $\alpha \rightarrow 0$, 
$f(\theta)\rightarrow \,\,$constant, and
\be
\label{eq:fpdm}
{\bf \Fpd} \rightarrow
{\bf \Fpd^{\rm m}} \equiv
- \gamma \pi a^2 \frac{\sH}{\sH + \cc} \bS
\int_0^{\pi} d\tp \sin\tp \cos\tp \Psi(\tp)~~~.
\ee
When $f(\theta) \neq f(\pi -\theta)$, the magnitude of ${\bf \Fpd}$ can be
much greater than $\Fpd^{\rm m}$, depending on the value of $C$.  For example,
when $\alpha \ll 1$ and $\alpha C \gg 1$, $\Fpd \approx \alpha C
\Fpd^{\rm m} / (\sH + \cc)$, for non-flipping grains.  
However, for realistic parameter choices and interstellar
conditions, it is usually the case that $\Fpd \le \Fpd^{\rm m}$.
Also note that, for non-flipping grains with $\tr=0$, $\Fpd =0$ when 
$\alpha C = \sH + \cc$.  In this case, the contributions from the various
momentum transfer processes exactly balance, regardless of the functional form
of $\Psi(\tp)$; of course, such a parameter conspiracy is unlikely for real 
grains in the ISM.  

\subsection{Computational Results}

In Figure \ref{fig:fpd1} (lower panel), we display the photodesorption 
asymmetry integral that appears in equation (\ref{eq:fpdm}) for 
$\Fpd^{\rm m}$, for carbonaceous and 
silicate grains.  This integral takes nearly the same values for the ISRF
as for a blackbody (cut off at $13.6 \eV$) with $\Tc = 3 \times 10^4 \K$.
We found
in Figures \ref{fig:asymmetry_factor_C} and \ref{fig:asymmetry_factor_sil}
that, for some grain sizes and photon energies, the electric intensity 
just below the grain surface
is unexpectedly concentrated in the hemisphere facing away from the 
radiation.  The same effect occurs with the external electric intensity, 
and leads to a ``backwards'' photodesorption force, for a range of grain 
sizes.  

In the top panel of Figure \ref{fig:fpd1}, we plot $\Fpd^{\rm m}/
\Frad$ for our canonical parameter values, given in Table 
\ref{tab:pdpars}.  The surface site area of $l^2 = 6 \Angstrom^2$ would result
if a site is a hexagon in a graphite sheet with C--C 
spacings of $\approx 1.5 \Angstrom$.  We assume that photodesorbed H atoms 
carry kinetic energy $E_{\rm H} = 2 \eV$ and newly-formed molecules carry
kinetic energy $E_{{\rm H}_2} = 1 \eV$, implying $\pH = \pHt$ (note that
$\pHt$ is not relevant for the calculation of $\Fpd^{\rm m}$, though).  
For large grains the photodesorption force dominates the 
radiation pressure for the blackbody spectrum, but is not nearly so 
important for the ISRF.  

\begin{figure}
\epsscale{1.00}
\plotone{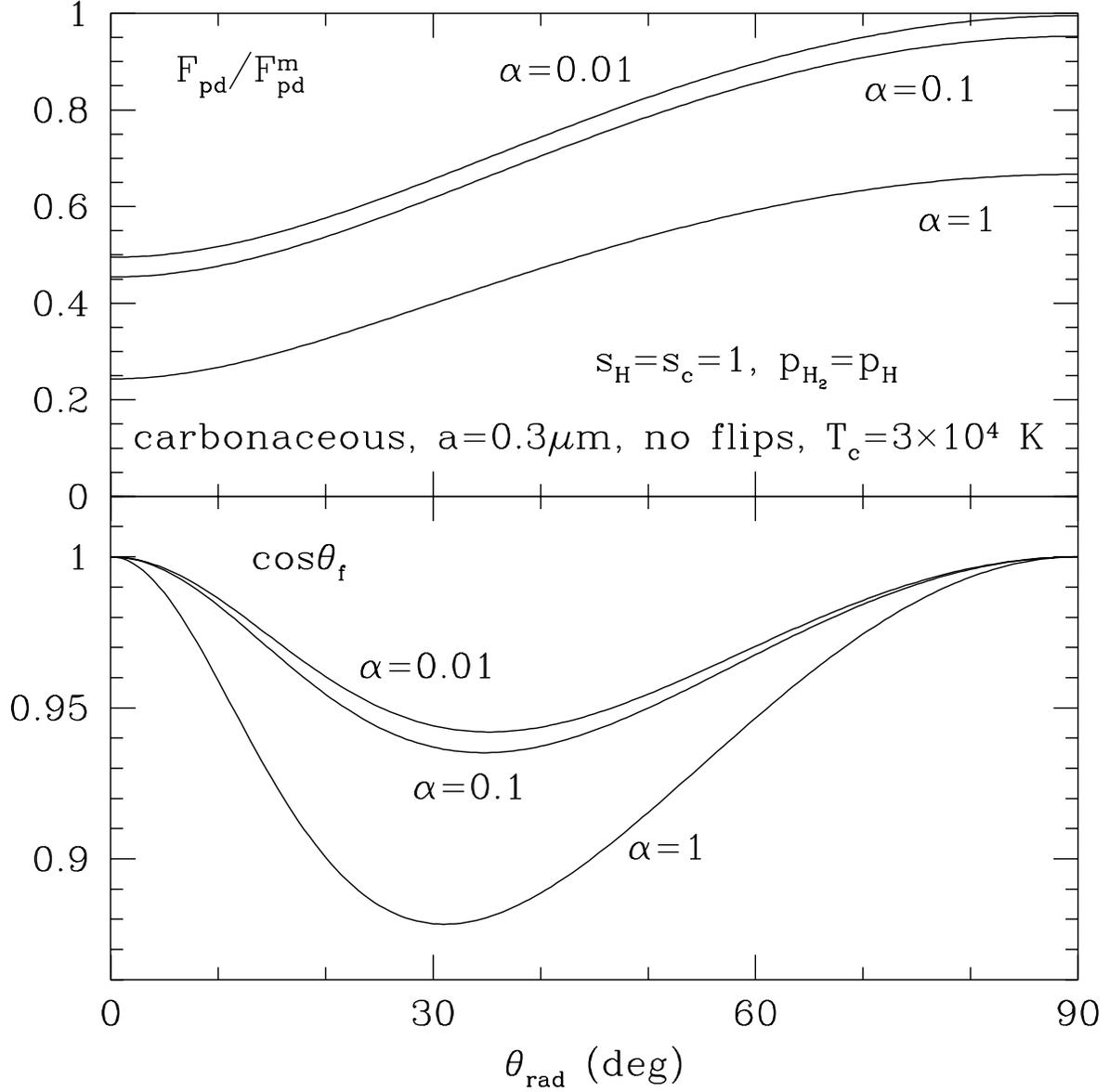}
\caption{
\label{fig:fpd2}
Upper panel:
The photodesorption force as a function of angle $\tr$ between $\hat{\omega}$
and ${\rm \hat{S}}$, normalized to 
${\bf F}_{pd}^m$ [see eq.\ (\ref{eq:fpdm})].
Lower panel:  $\theta_{\rm f}$ is the angle between
${\bf \Fpd}$ and ${\bf \hat{S}}$.	
        }
\end{figure}

In the upper panel of
Figure \ref{fig:fpd2} we display $\Fpd / \Fpd^{\rm m}$ for non-flipping
carbonaceous
grains with $a=0.3 \micron$, for a few values of $\alpha$.\footnote{
For our assumed carbonaceous grain
parameter values, $G/n_{\rm H} \approx 4.4 \times
10^{-3} \alpha \sqrt{T} \K^{-1/2} \cm^3$.}  We also
assume that $\sH=\cc=1$, $\pHt=\pH$, and a blackbody radiation spectrum
(cut off at $13.6\eV$) with $\Tc=3 \times 10^4 \K$.  In the lower panel of
Figure \ref{fig:fpd2} we display $\cos \theta_{\rm f}$, where $\theta_{\rm f}$
is the angle between ${\bf \Fpd}$ and $\bS$.  The results for
other grain sizes for which $\Fpd$ is significant, for silicates, and for
illumination by the ISRF are very similar.  

\begin{figure}
\epsscale{1.00}
\plotone{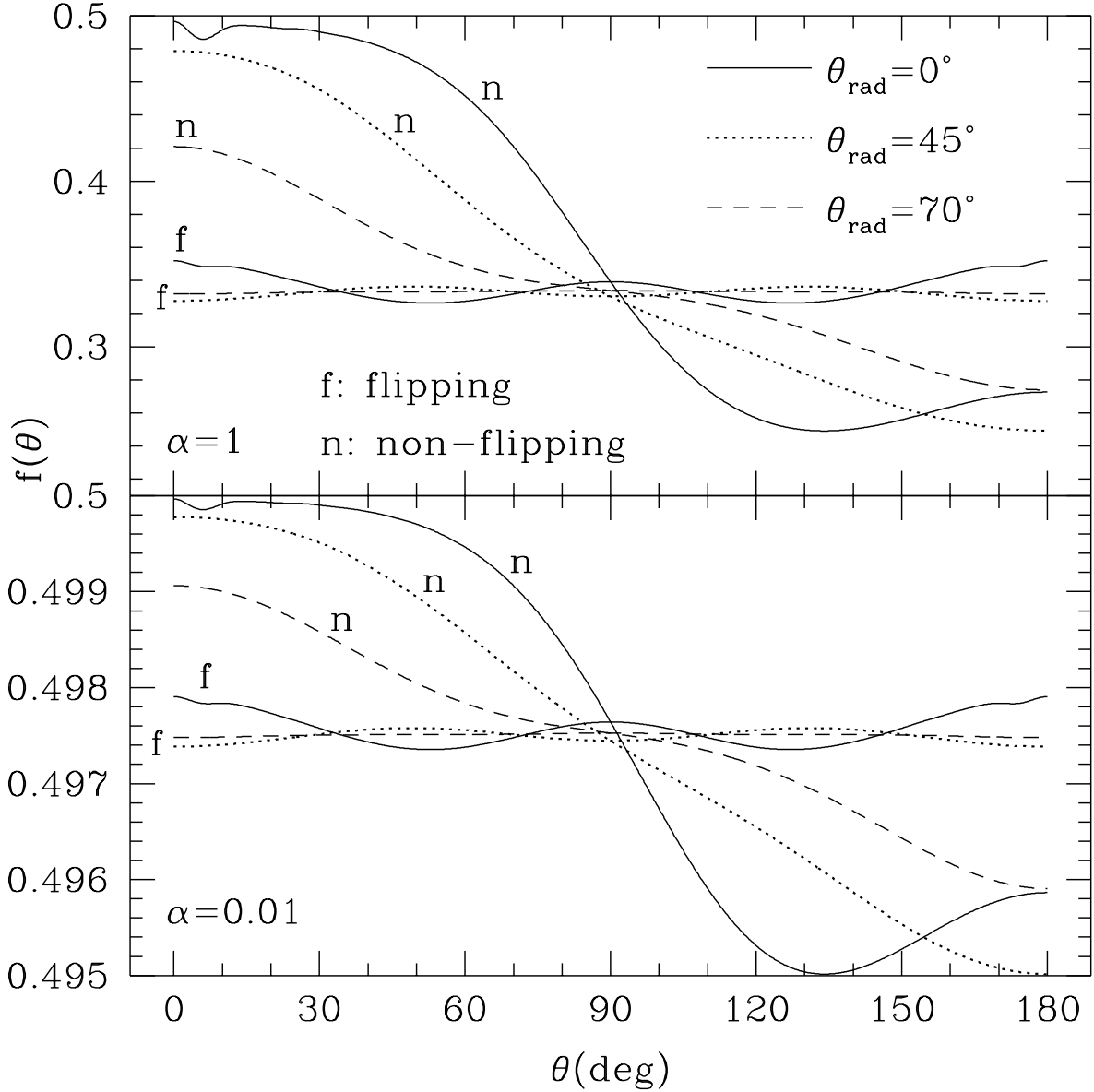}
\caption{
\label{fig:occfrac}
The occupation fraction $f$ as a function of the angle $\theta$ with respect
to the spin axis, for carbonaceous grains with $a=0.3\micron$,
$\sH=\cc=1$, $\pHt = \pH$,  
a dilute blackbody radiation field with $\Tc = 3 \times 10^4 \K$,
and two values of $\alpha$: $\alpha=0.01$ and $\alpha=1$.
        }
\end{figure}

As $\tr$ increases, H atoms which stick to sites on the ``dark''
side of the grain have greater opportunity to photodesorb from the side
facing the radiation, as they are transported there by the rotation; thus 
the photodesorption force increases with $\tr$.  For flipping grains, the 
occupation fraction $f$ depends only mildly on $\theta$ (Figure 
\ref{fig:occfrac}), so that $\Fpd 
\approx \Fpd^{\rm m}$ and $\cos \theta_{\rm f} \approx 1$ for all values of
$\tr$.  

\begin{figure}
\epsscale{1.00}
\plotone{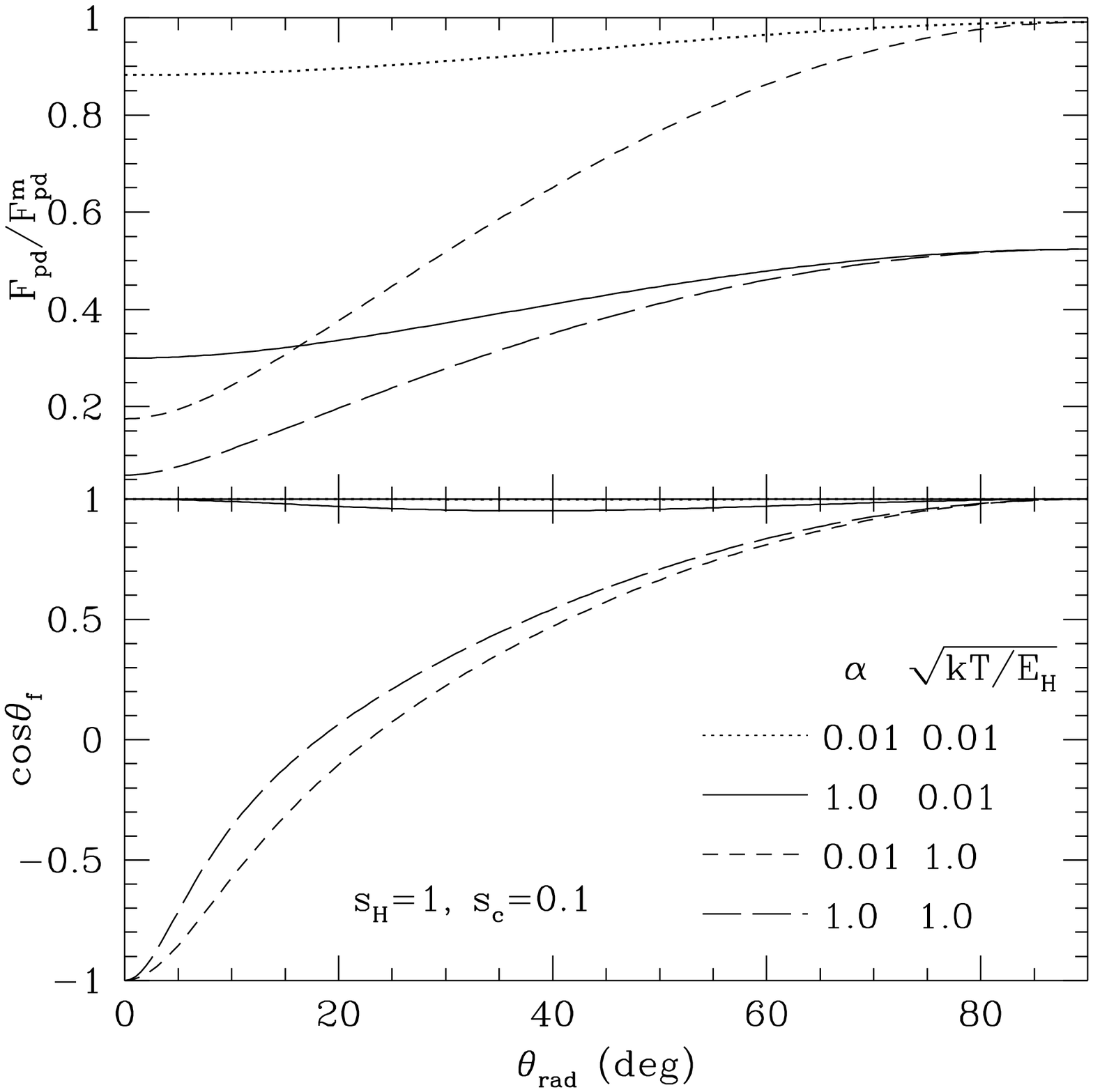}
\caption{
\label{fig:fpd3}
$\Fpd / \Fpd^{\rm m}$ and $\cos \theta_{\rm f}$ for non-flipping carbonaceous 
grains with $a=0.3 \micron$, $\sH=1$, $\cc=0.1$, $\pHt = \pH$, a blackbody
radiation field with $\Tc=3 \times 10^4 \K$, and various values of 
($\alpha$, $\sqrt{kT/E_{\rm H}}$).
        }
\end{figure}

In Figure \ref{fig:fpd3} we plot $\Fpd / \Fpd^{\rm m}$ and 
$\cos \theta_{\rm f}$ for $\cc=0.1$ and various combinations of 
$\alpha$ and $(k T/E_{\rm H})^{1/2}$;
otherwise the parameters are the same as for Figure \ref{fig:fpd2}.  
When $\alpha = (k T/E_{\rm H})^{1/2} = 0.01$, $\cos \theta_{\rm f}$ is very
nearly 1 for all values of $\tr$, much more so than when $\cc=1$ and 
$\alpha=0.01$.  Although $f(\theta)$ is somewhat more
constant for the case with $\sH=1$, $C$ is also substantially larger,
and it accounts for the bulk of the component perpendicular to 
$\bS$.  When $(k T/E_{\rm H})^{1/2} = 1$, $C$ is large enough 
that $\cos \theta_{\rm f} \approx -1$ for $\tr = 0$.

\section{Drag Forces \label{sec:drag}}

Baines et al. (1965) calculated the drag force on a
spherical particle drifting through a dilute gas, assuming that 
colliding atoms reflect elastically.  Draine \& Salpeter 
(1979) obtained an analytical expression which accurately reproduces the 
Baines et al. numerical results, and this is widely employed in studies
of drifting interstellar grains.  For subsonic drift speeds, the drag due to 
collisions with neutral atoms of element $j$, in the absence of surface 
processes, is given by
\be
\label{eq:fdragh0}
F_{{\rm drag},j}^0 \approx \pi a^2 \  \frac{8}{3} n_j v_{\rm d}
\left(\frac{2 m_p k T}{\pi} \right)^{1/2} A_j^{1/2}~~~,
\ee
where $n_j$ and $A_j$ are the number density and mass number of element $j$, 
respectively.  

The Coulomb drag, due to interactions with distant ions,
is often important in interstellar applications.  For subsonic drift 
speeds, the Coulomb drag is given by
\be
\Fdc \approx F_{\rm drag,H}^0
\times \frac{1}{2} \sum_i x_i \left( \frac{m_i}{m_p}
\right)^{1/2} \sum_Z f_Z \left( \frac{Z e^2}{a k T}\right)^2 
\ln \left[ \frac{3 (kT)^{3/2}}{2 e^3 |Z| {(\pi x \nH)}^{1/2}} \right]
\ee
(Draine \& Salpeter 1979). The sum is over ions $i$, and $x = \sum_i x_i$.
Since the grain charge fluctuates on a time scale short compared to the
drag time, we average over charge states. 
The dominant ions in neutral gas are C$^+$ and H$^+$.
We take the gas-phase C abundance
to be ${\rm C}/{\rm H} = 1.4 \times 10^{-4}$ (Cardelli et al.~1996;
Sofia et al.~1997).  Thus, when calculating $\Fdc$,
we assume that $x_{\rm C} = x$, $x_{\rm H}\approx0$
when $x \le 1.4 \times 10^{-4}$, and 
$x_{\rm C} = 1.4 \times 10^{-4}$, $x_{\rm H} = x - 1.4 \times 10^{-4}$
when $x > 1.4 \times 10^{-4}$.

Next, we consider how the surface processes
considered in the previous section, namely H$_2$ formation and 
photodesorption, affect the drag force ${\bf \Fdh}$
due to collisions with atomic H.

For a drifting grain, the arrival rate $\Rarr$ depends on the angle
$\tdp$ with respect to the grain's velocity ${\bf u}$:
\be
\Rarr (\tdp) = \Rao D(\tdp)~~~,
\ee
where 
\be
D(\tdp) = \exp \left( - y^2 \right) + \sqrt{\pi} y \left[ 1 + \frac{2
}{\sqrt{\pi}} \int_0^y dt \exp \left( - t^2 \right) \right]~~~,
\ee
with $y \equiv \sqrt{\beta} u \cos \tdp$.  The expansion 
\be
D(\tdp) \approx 1+\sqrt{\pi} y + y^2 - \frac{1}{6} y^4
\ee
is accurate to within $1 \%$ for $-0.653 \le y \le 1.05$.

The momentum transfer rates due to reflections and sticking are modified 
from the non-drifting case by $P(v) v^2 \rightarrow P(v)(v+u\cos\tdp)^2$ 
in Table 
\ref{tab:momtransfer} (and the lower integrand is now $-u \cos\tdp$).  
The rates obtained for non-drifting grains must be multiplied by the factor
$2 E(\tdp)/\sqrt{\pi}$, where
\begin{eqnarray}
E(\tdp) & = & 
y \exp (- y^2) + (1 + 2 y^2) \left( 
\frac{\sqrt{\pi}}{2} + \int_0^y \exp(- t^2) dt \right) \nonumber\\
& \approx &
\frac{\sqrt{\pi}}{2} + 2 y + \sqrt{\pi} y^2 + \frac{2}{3} y^3 - \frac{1}
{15} y^5~~~.
\label{eq:exp_e}
\end{eqnarray}
The expansion in equation (\ref{eq:exp_e}) is accurate to within $1 \%$ for
$-0.749 \le y \le 1.43$.


Photodesorption substantially complicates the drag, since the occupation 
fraction is no longer constant, but is given by
\be
\label{eq:fu}
f(\theta) = \sH \left[ \sH+\cc + \alpha \frac{\bar{\Psi}(\theta)}
{\bar{D}(\theta)} \right]^{-1}~~~,
\ee
for non-flipping grains. Of course, for flipping grains 
$\bar{\Psi}(\theta)/\bar{D}(\theta)$ is 
replaced by $[\bar{\Psi}(\theta)+\bar{\Psi}(\pi - \theta)]/[\bar{D}(\theta)
+\bar{D}(\pi - \theta)]$ in equation (\ref{eq:fu}).  
The drag force can be found by evaluating the 
total force resulting from collisions of H with the grain
(i.e. ${\bf \Fpd} + {\bf \Fdh}$) and subtracting ${\bf \Fpd}$;
the total force is given by
\begin{eqnarray}
{\bf \Fpd} + {\bf \Fdh} = - \frac{a^2}{l^2} \int_0^{\pi} d\theta \sin
\theta \int_0^{2 \pi} d\phi \nonumber\\
\left\{ \frac{2}{3} \Ro \pH f(\theta) \Psi
(\tp) + \frac{2}{3} \Rao \cc \pHt f(\theta) D(\tdp) + \left[ 2 -\sH +
f(\theta) (\sH-\cc) \right] \frac{m_p}{\sqrt{\beta}} \Rao E(\tdp) \right\}
\nonumber\\
\left[ \bz \cos\theta + \bx \sin\theta \cos\phi + \by \sin\theta
\sin\phi \right]~~~.
\label{eq:fpd+fdragh}
\end{eqnarray}

To first order in $u$, equation (\ref{eq:fpd+fdragh}) yields, for flipping
grains,
\be
\label{eq:fdragh1}
\frac{{\bf \Fdh}}{\Fdh^0} = -\frac{2-\sH}{2} {\bf \hat{u}} -
\frac{3}{8} \left[ \sH - \cc +\frac{\cc \pHt}{3} 
\left( \frac{\pi}{2 m_p kT}\right)^{1/2} 
\right] \int_0^{\pi} d\theta \sin\theta f_0(\theta)
\left[\sin^2 \theta {\bf \hat{u}} - (1-3\cos^2 \theta) \cos\tu \bomega
\right]~~~,
\ee
where $f_0(\theta)$ is the occupation fraction that would hold for the same
grain if $u=0$.
For non-flipping grains, the following term must be added to the right-hand
side of equation (\ref{eq:fdragh1}):
\be
- \frac{\alpha \cos\tu}{4 \sH} \int_0^{\pi} d\theta \sin\theta \cos\theta
f_0^2(\theta) \bar{\Psi}(\theta) \left[ {\bf J_1}(\theta) + 
{\bf J_2}(\theta)\right]~~~,
\ee
where
\be
{\bf J_1}(\theta) = \bz \cos\theta \left[ \frac{3\pi}{4}(\sH - \cc) + 
\cc \pHt \left( \frac{\pi}{2 m_p k T}\right)^{1/2} \right]
\ee
and
\be
{\bf J_2}(\theta) = \alpha \pH \left( \frac{\pi}{2 m_p k T}\right)^{1/2}
\left[ \bz \cos\theta \bar{\Psi}(\theta) + \bx \sin\theta \frac{1}{2\pi}
\int_0^{\pi} d\phi \cos\phi \Psi(\tp) \right]~~~.
\ee
The above approximation works best for small $a$ and $\alpha$.  For the 
range of conditions $T \in (50, 6000) \K$ and $\sqrt{\beta} u \le 0.1$,
it is valid to within $1\%$ when $a\le 0.1\micron$, $\alpha \le 1$;
$3\%$ when $a\le 1\micron$, $\alpha \le 1$; $10\%$ when 
$a\le 0.1\micron$, $\alpha \le 10$; and $30\%$ when 
$a\le 1\micron$, $\alpha \le 10$.  
The direction of ${\bf \Fdh}$ can differ
substantially from $- {\bf u}$ when $\alpha \ge 1$, depending on the relative
orientations of ${\bf u}$, $\bomega$, and $\bS$.  

\section{Starlight Anisotropy \label{sec:anisotropy}}

Here we
estimate the radiation field anisotropy in the diffuse ISM, by considering
the solar neighborhood to be typical and adding up the flux density over 
the entire sky.  Most of the flux in the visible and UV comes directly
from stars; starlight scattered off of dust grains (the diffuse
galactic light, DGL) makes a non-negligible additional
contribution (see, e.g., Witt 1989).
Ideally, we would like to have a catalog of stars with accurate photometry
and complete down to the magnitude beyond which fainter stars do not 
significantly contribute, along with accurate photometry of the DGL.  

\begin{figure}
\epsscale{1.00}
\plotone{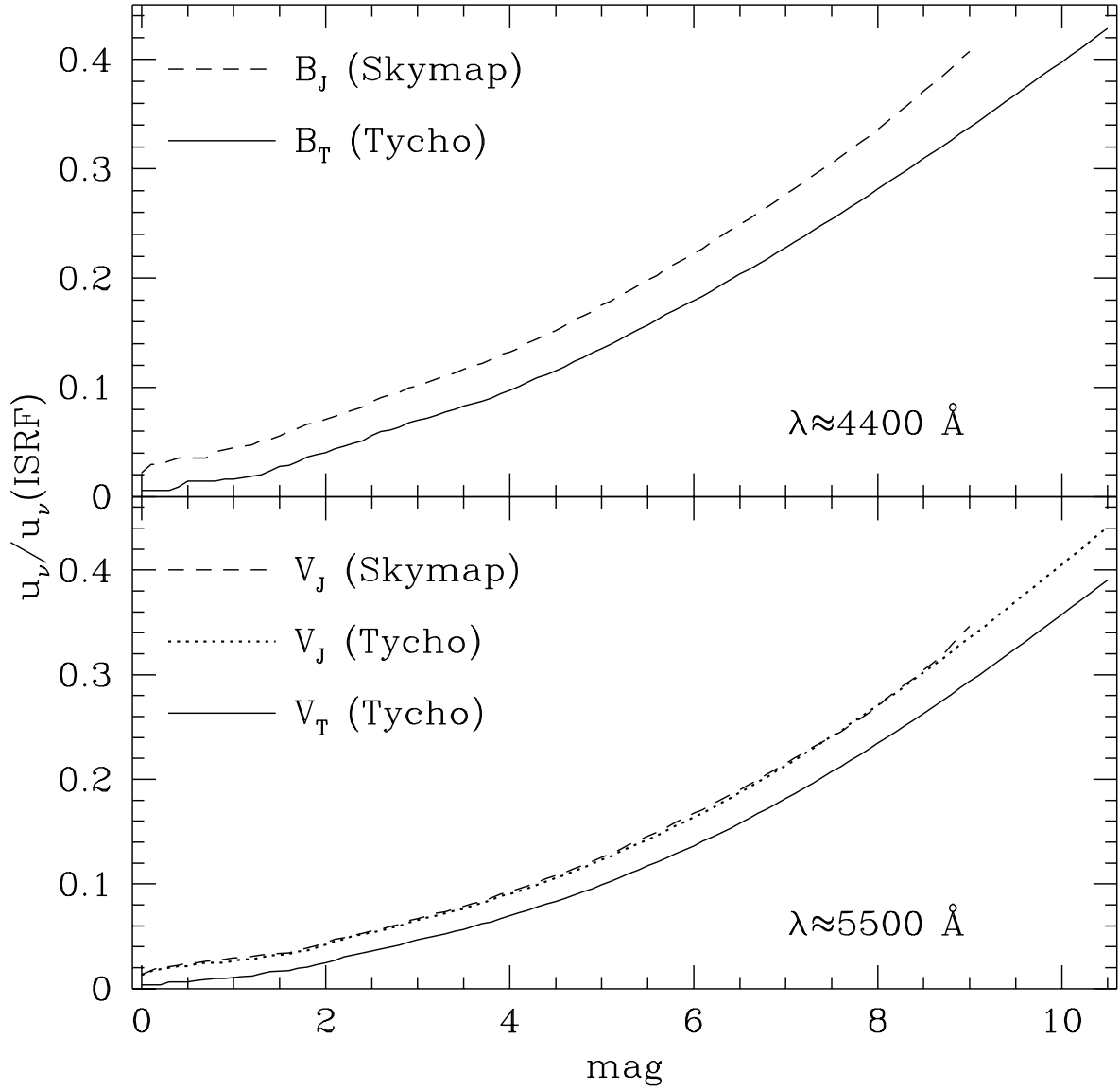}
\caption{
\label{fig:flambda}
        Energy density per wavelength interval, normalized to that of the 
ISRF, as a function of limiting magnitude.  Results are from the Skymap
and Tycho star catalogs.  The Tycho fluxes are systematically low due to
the absence of photometry for thousands of bright stars.
        }
\end{figure}

Although this ideal is not currently available, there are a few catalogs
from which useful estimates can be made.  The Skymap Star Catalog, 
Version 3.7 (Slater \& Hashmall 1992) is claimed to be complete down to 
magnitude 9 in the $B$ and $V$ bands, and the Tycho Catalog (ESA 1997) is 
claimed to be $99.9 \%$ complete down to $V \sim 10.5$.  However, the
photometry is missing for thousands of stars in the Tycho Catalog, 
including many bright stars.  In Figure \ref{fig:flambda} we plot the 
integrated energy densities per unit frequency $u_{\nu}$, normalized to the 
ISRF, as a function of the limiting stellar
magnitude.  The Skymap results are for 
the Johnson $B$ and $V$ bands ($B_{\rm J}$ and $V_{\rm J}$ respectively);
the Tycho bands $B_{\rm T}$ and $V_{\rm T}$
closely resemble the corresponding Johnson bands.  The Tycho catalog also
gives indirectly-derived values for $V_{\rm J}$ for all stars, including
those for which Tycho photometry is missing.  The $B_{\rm T}$ and 
$V_{\rm T}$ energy densities are systematically lower than the 
$B_{\rm J}$ and $V_{\rm J}$ energy densities due to omission of stars
lacking Tycho photometry.  
The $V_{\rm J}$ fluxes from the two catalogs agree very well, except at
the limit of the Skymap catalog.  Note that at magnitude 10
the curves are still concave up, with
the integrated starlight accounting for less than half of the ISRF.

\begin{figure}
\epsscale{1.00}
\plotone{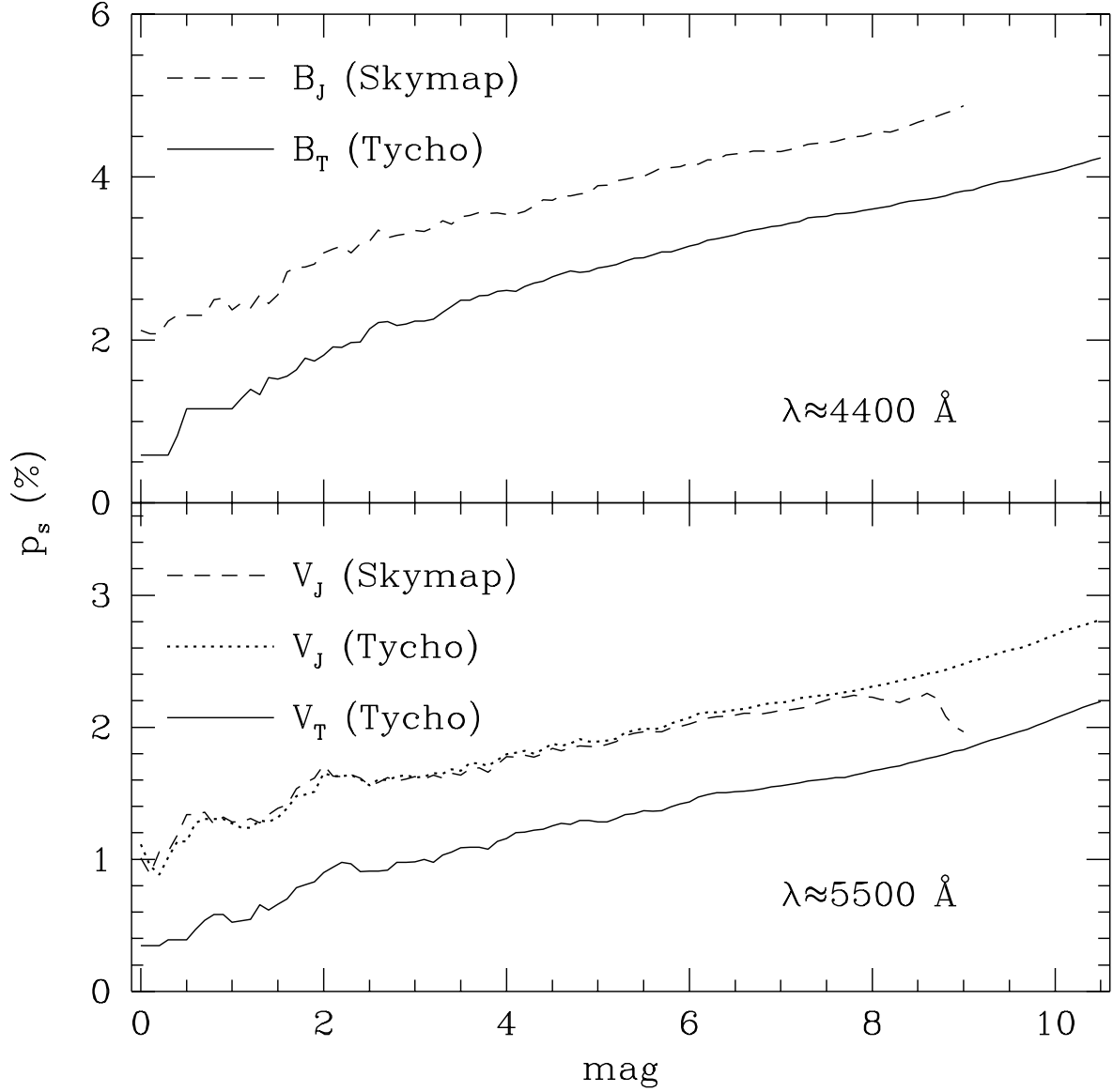}
\caption{
\label{fig:dipole}
        The normalized dipole moment $p_s$, 
i.e., the fraction of the energy density in the ISRF that
is in the anisotropic component, as a function of limiting magnitude.  The
Tycho anisotropies are systematically low due to the absence of photometry 
for many of the brightest stars.
        }
\end{figure}

The normalized starlight dipole moment is given by 
\be
{\bf p_s} = \frac{\sum_s u_{\nu}(s) {\bf \hat{n}}(s)}  
{u_{\nu}({\rm ISRF})},
\ee
where $u_{\nu}(s)$ is the energy density per frequency interval due to
star $s$ and 
${\bf \hat{n}}(s)$ is the direction to star $s$.  
In Figure \ref{fig:dipole} we plot $p_s$.
Again, the $B_{\rm T}$ and $V_{\rm T}$ results 
are offset, because of the absence of photometry for bright stars.  The
abrupt decrease in the V band normalized
dipole moment at the faint end of the 
Skymap catalog is probably incorrect.  It is unlikely that the fainter 
stars and DGL would conspire to produce an anisotropy directed opposite
to that resulting from the brighter stars; thus it appears that the 
anisotropy in $B$ ($V$) is at least $5 \%$ ($3 \%$).  For $B$, the 
(right ascension, declination) of the anisotropy remain quite constant
for limiting magnitudes of 4 through 9, with value ($115^{\circ}$, 
$-43^{\circ}$).  For $V$, the anisotropy direction gradually shifts to
($115^{\circ}$, $-51^{\circ}$) at magnitude $10.5$.  

The best available catalog for studying the anisotropy in the UV was 
constructed by the S2/68 experiment (Boksenberg et al. 1973) on the 
European Space Research Organization's TD-1 satellite, which provided 
broadband flux measurements in four channels.  The full catalog
was not published, but Gondhalekar, Phillips, \& Wilson (1980) presented 
tables of the integrated starlight in hundreds of patches, covering the 
entire sky.  Gondhalekar et al. also estimated the flux contribution from the 
DGL and from stars too faint to be included in the catalog, using background
data.  In some cases, two or more stars were observed simultaneously.  Such 
blended stars were excluded from the catalog, and Gondhalekar et al. 
estimated their contribution to the total flux, although the
distribution of these unresolved stars
on the sky was not reported.  The Mathis et al. (1983) ISRF was 
constructed to maximize consistency with the Gondhalekar et al. (1980)
results and with other, limited-coverage UV surveys. 
The adopted ISRF fluxes are somewhat higher than the total fluxes of 
Gondhalekar et al.  

In estimating the dipole in the UV, we use tables of integrated starlight
in $10^{\circ} \times 10^{\circ}$ patches in Gondhalekar (1989).  These 
differ somewhat from the fluxes in Gondhalekar et al. (1980), because the
absolute calibration of the S2/68 experiment was revised.  In Table 
\ref{tab:uv} we give $\nu u_{\nu}$ and the normalized dipole moment $p_s$
for the four UV bands; the listed dipole moments are
likely lower limits.  Note that the normalized dipole 
moment is substantially greater in the UV than in the visible. 
This difference presumably results from the differences in the
stellar populations giving rise to the two spectral regions and the greater
importance of extinction by dust in the UV.  The UV radiation
is dominated by early-type stars, which are relatively few in number and 
hence less evenly distributed than main sequence stars in general, all of 
which contribute to the visible radiation.  In Table \ref{tab:dir} we 
display the direction of the anisotropy for all of the considered bands.

For simplicity, we will assume that the anisotropy is independent of 
wavelength when calculating drift speeds in the following section.
We will adopt an anisotropy of
$10 \%$, intermediate between the anisotropies inferred for the visible and
the UV.

\section{Gas-Grain Drift in the Diffuse ISM \label{sec:drift_diffuse}}

We will estimate drift speeds for grains in the cold neutral 
medium (CNM; $T=100\K$, $\nH=30\cm^{-3}$, $n_e=0.045\cm^{-3}$) and the warm 
neutral medium (WNM; $T=6000\K$, $\nH=0.3\cm^{-3}$, $n_e=0.03\cm^{-3}$).

In the presence of a magnetic field ${\bf B}$, the steady state drift 
velocity is given by
\be
\left(v_{\parallel},v_{\perp},v_{\perp\,\perp}\right)
 = \frac{F}{\kappa}
\left(  \cos\theta ,
        \frac{\sin\theta}{1+(\omega\tau)^2} ,
        \frac{\omega\tau \sin\theta}{1+(\omega\tau)^2}
\right)
\label{eq:vector_vd}
\ee
where ${\bf F}$ is the net force excluding the magnetic and drag forces,
${\bf F}_{\rm drag} = - \kappa {\bf v_d}$, 
$\theta$ is the angle between ${\bf B}$ and ${\bf F}$, 
$\omega\tau= Z e B/c\kappa$ is the product of the gyrofrequency
and gas-drag time, and
the subscripts ($\parallel$, $\perp$, $\perp \, \perp$) denote 
components (parallel to ${\bf B}$, perpendicular to 
${\bf B}$ and in the plane spanned by ${\bf B}$ and ${\bf F}$, 
perpendicular to the plane spanned by ${\bf B}$ and ${\bf F}$).  
Thus, drift across magnetic field lines is suppressed by a factor
$[1 + (\omega \tau)^2]^{1/2}$.

\begin{figure}
\epsscale{1.00}
\plotone{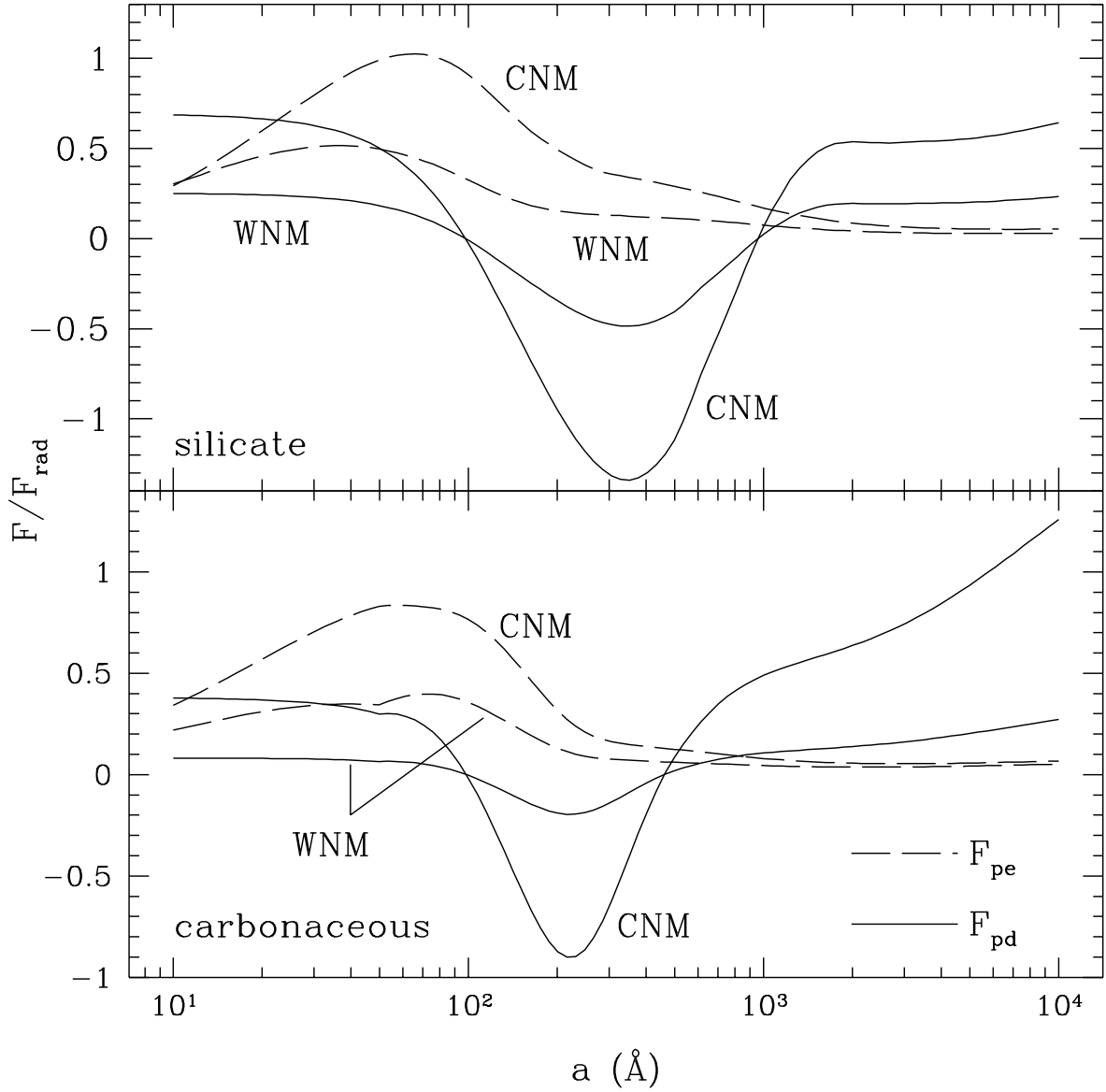}
\caption{
\label{fig:force_diffuse}
The photoelectric and photodesorption forces for grains in
the warm and cold neutral media.
        }
\end{figure}

\begin{figure}
\epsscale{1.00}
\plotone{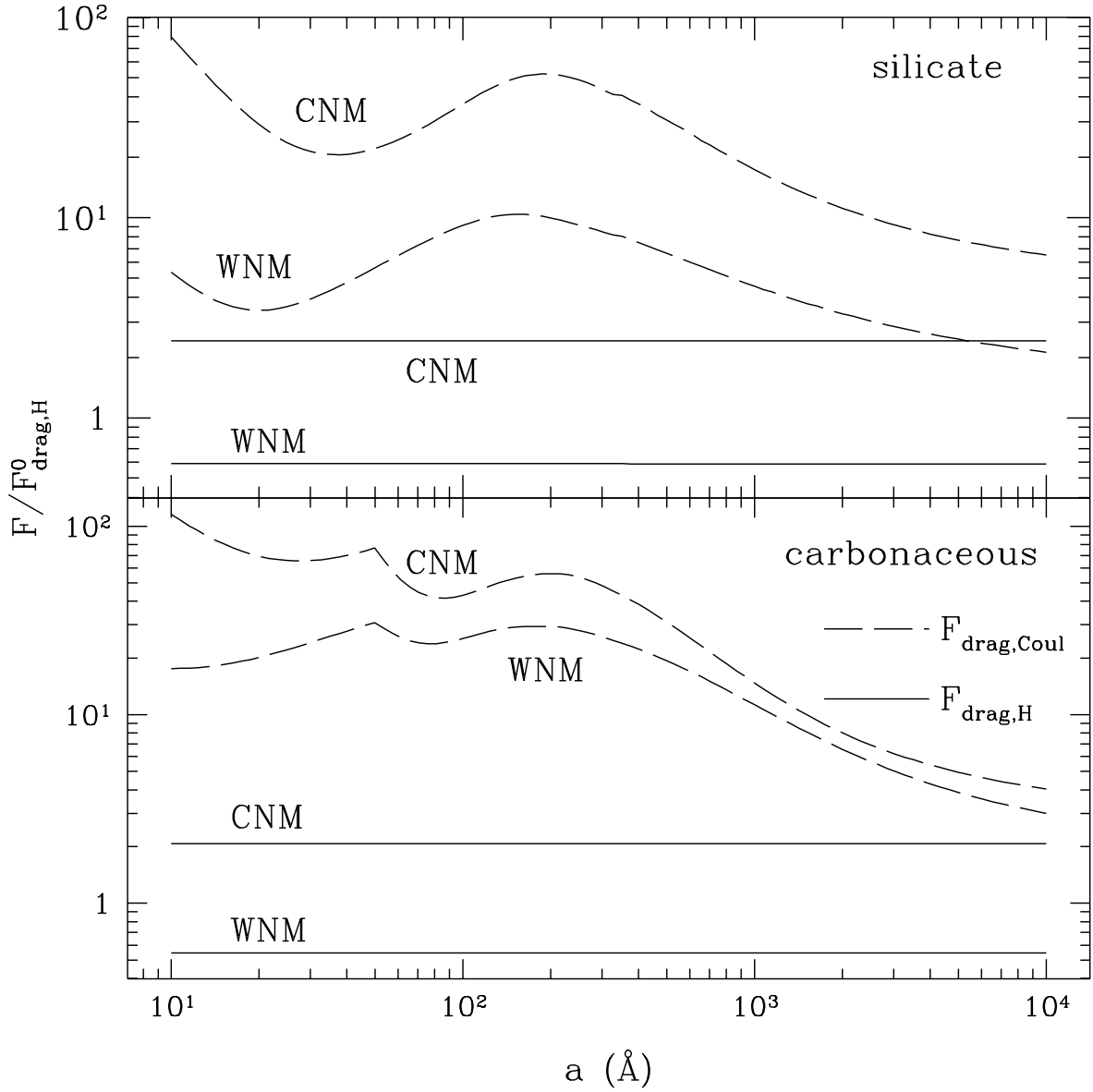}
\caption{
\label{fig:drag_diffuse}
The drag forces due to collisions with neutral H and to Coulomb interactions
with distant ions.
        }
\end{figure}

\begin{figure}
\epsscale{1.00}
\plotone{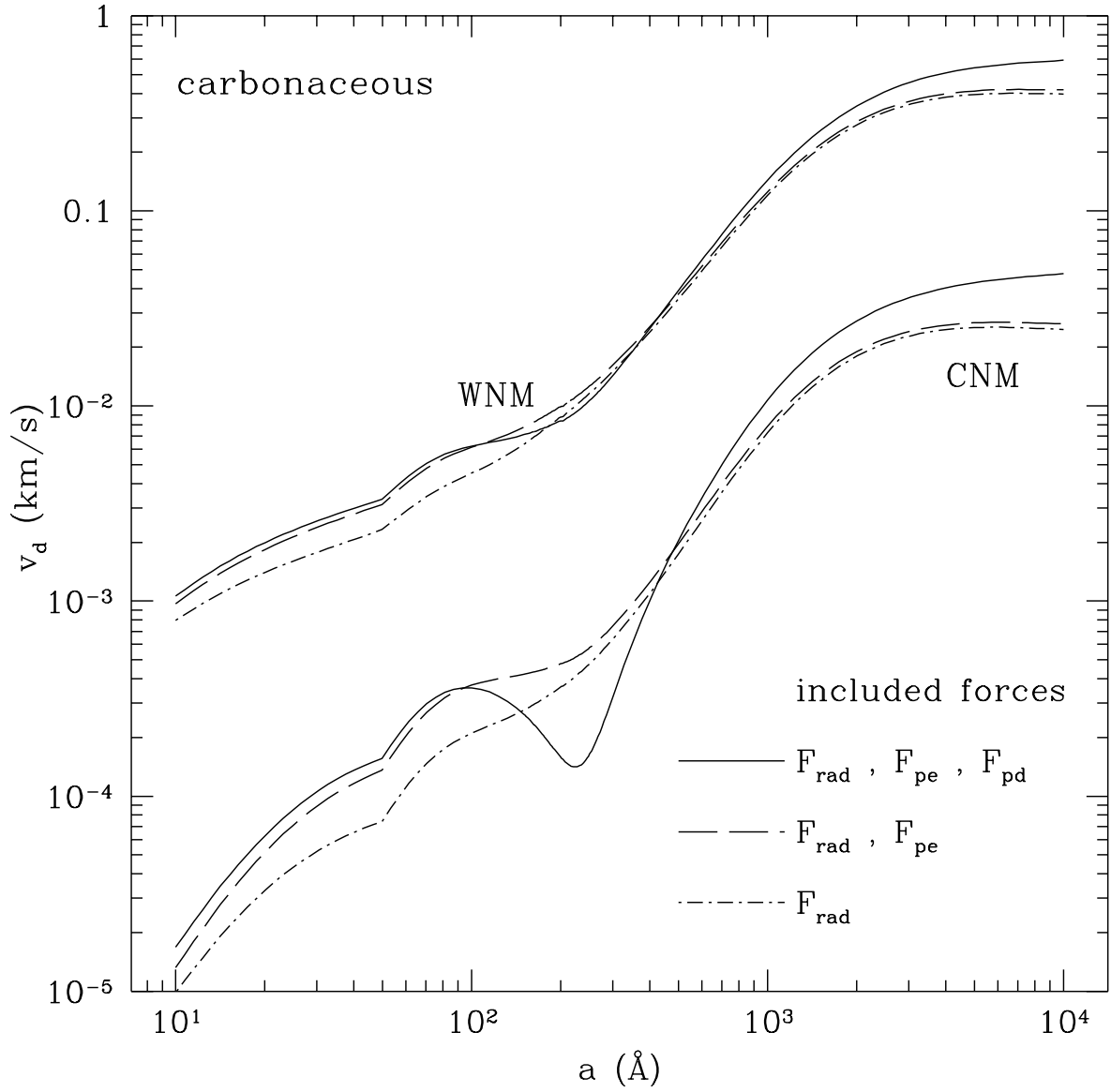}
\caption{
\label{fig:vdrift_gra}
The drift speed for carbonaceous grains in the warm and cold neutral media.
        }
\end{figure}

\begin{figure}
\epsscale{1.00}
\plotone{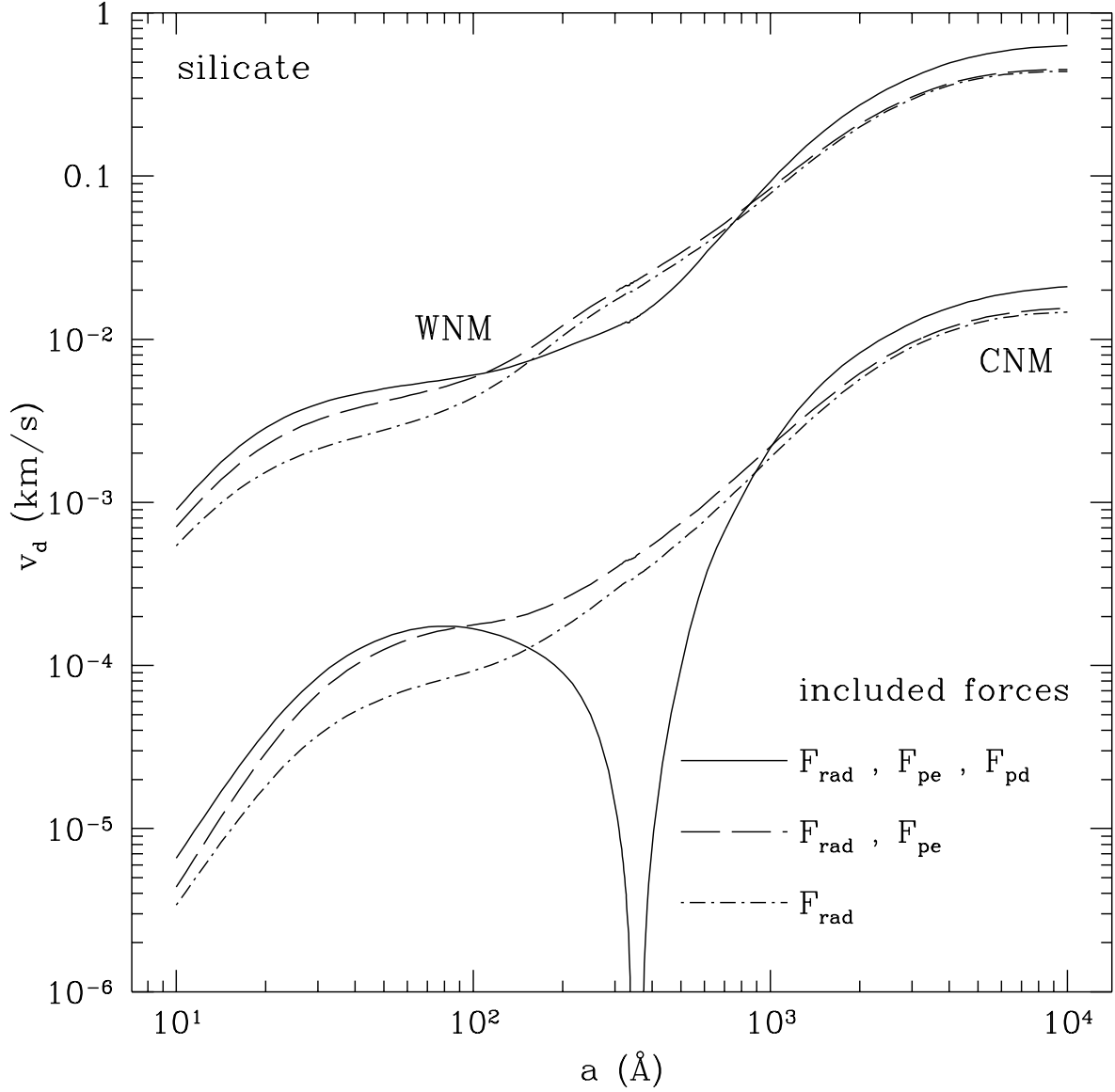}
\caption{
\label{fig:vdrift_sil}
The drift speed for silicate grains in the warm and cold neutral media.
Coincidentally, the minimum value of $\Frad + \Fpe + \Fpd$ in the CNM is
very close to zero; hence the sharp drop in $v_{\rm d}$ when 
$a \approx 350 \Angstrom$.
        }
\end{figure}

\begin{figure}
\epsscale{1.00}
\plotone{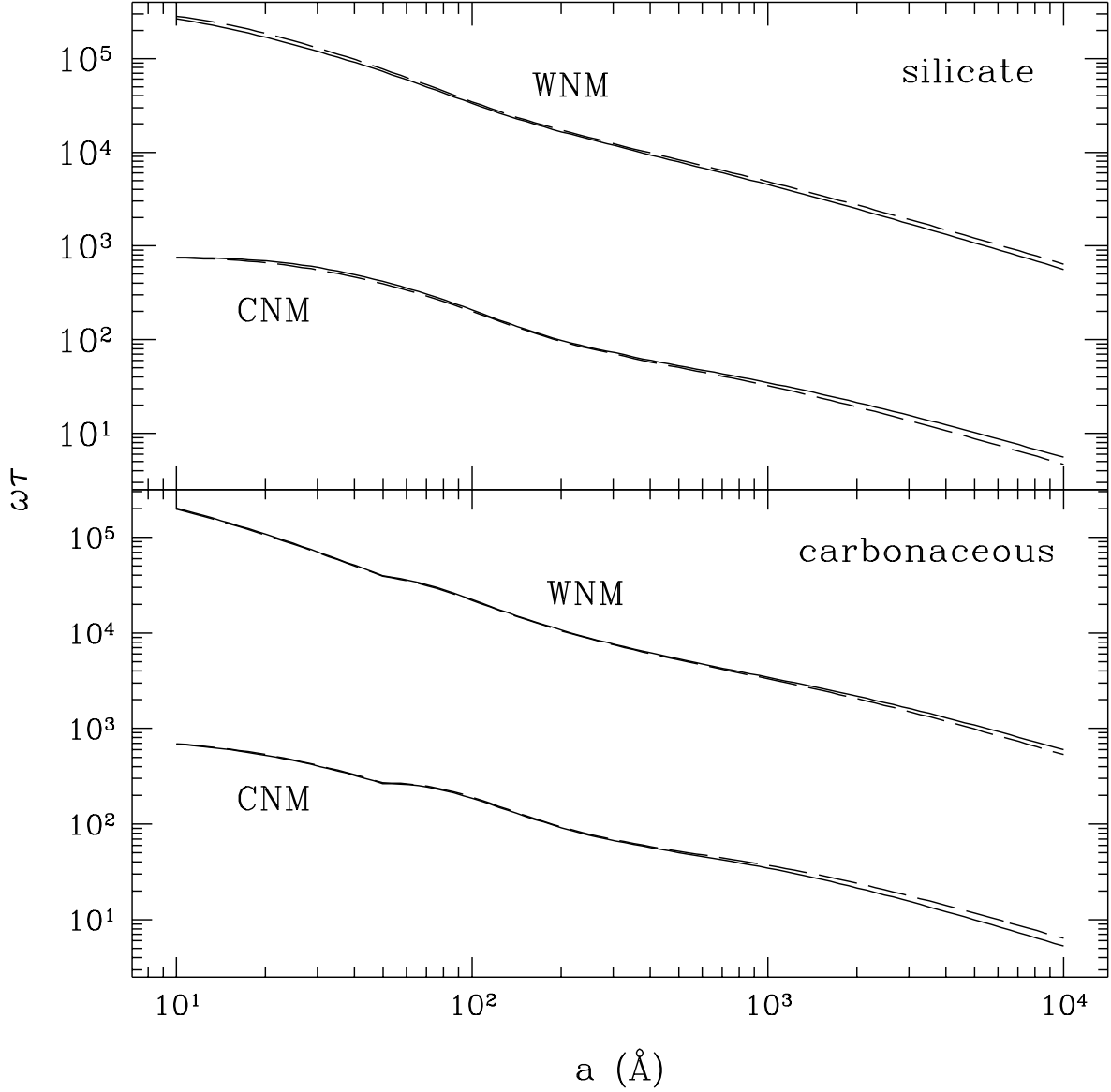}
\caption{
\label{fig:omegatau}
	The product $\omega \tau$, where $\omega$ is the gyrofrequency for the
	charged grains in the ambient magnetic field, 
	and $\tau$ is the drag time.
	Drift transverse to the magnetic field is suppressed by a factor
	$[1+(\omega\tau)^2]^{1/2}$
	[see eq.\ (\ref{eq:vector_vd})].
        }
\end{figure}

We display $\Fpe$ and $\Fpd$ for the CNM and WNM in Figure 
\ref{fig:force_diffuse}, and $\Fdh$ and $\Fdc$ in Figure
\ref{fig:drag_diffuse}.  In order to estimate
the largest effect that photodesorption is likely to have, we adopt
$\tr = 90^{\circ}$ (i.e.~spin axis perpendicular to the net radiation flux);
in this case ${\bf \Fpd} \parallel {\bf \hat{S}}$ and ${\bf \Fdh} \parallel
- {\bf \hat{S}}$.  In order to gauge the importance of the photoelectric
and photodesorption forces, we compute drift speeds both with and without
including them.  Specifically, we consider the following three cases:
\be
{\rm Case \ I}\ \ \ \ \ \Frad = 
F_{\rm drag,H}^0 
+ \Fdhe + \Fdc~~~,
\ee
\be
{\rm Case \ II}\ \ \ \ \ \Frad + \Fpe = 
F_{\rm drag,H}^0 
+ \Fdhe + \Fdc~~~,
\ee
and
\be
{\rm Case \ III}\ \ \ \ \ \Frad + \Fpe + \Fpd = \Fdh + \Fdhe + \Fdc~~~.
\ee
The resulting drift speeds are plotted in Figures \ref{fig:vdrift_gra}
and \ref{fig:vdrift_sil}.  
For $a > 0.1\micron$ carbonaceous grains, we find drift speeds
$v_d > 0.01 \kms$ in the CNM, and 
$v_d > 0.15 \kms$ in the WNM;
for $a > 0.1\micron$ silicate grains, we find
$v_d > 0.002 \kms$ in the CNM, and $v_d > 0.1 \kms$ in the WNM.
The softness of the ISRF prevents $\Fpe$ and
$\Fpd$ from substantially increasing the drift speeds over those that would
hold if only $\Frad$ were acting.  These drift speeds were determined under
the assumption that the radiation flux ${\bf \hat{S}}$
is along the magnetic field.  As seen in Figure \ref{fig:omegatau}, the 
suppresion factor for drift perpendicular to the field 
is substantial even for micron-sized grains.

\subsection{Consequences}

Since grains of different sizes drift with different speeds, grains
collide, with coagulation likely at low collision
speeds.  As $a \rightarrow 0$, $v_{\rm d} \rightarrow 0$, so that if
grains always stick upon colliding, the timescale $t_a$ for a very small 
grain to become attached to another grain is given by
\be
t_a^{-1} \approx \pi \int_{a_{\rm min}}^{a_{\rm max}} a^2 v_{\rm d} (a)
\frac{d n_{\rm gr}}{da} da.
\ee
Here we estimate the maximum likely values for $t_a$ in the CNM and 
WNM, by adopting the drift speeds computed for case III above 
(i.e., including surface processes and assuming  
${\bf \hat{S}}\parallel{\bf B}$; $t_a$ is only slightly longer for 
cases I or II).  We adopt the grain size 
distribution from Weingartner \& Draine (2001a) with $R_V=3.1$ and 
$b_{\rm C}=6\times 10^{-5}$ (their favored distribution for
the average diffuse ISM).  For coagulation with the carbonaceous grain 
population, $t_a \approx 8 \times 10^{9}
\, {\rm yr}$ ($6 \times 10^{10} \, {\rm yr}$) in the CNM (WNM).  For
coagulation with the silicate population, $t_a \approx 1.5 \times 10^{10}
\, {\rm yr}$ ($4 \times 10^{10} \, {\rm yr}$) in the CNM (WNM).
Thus, drift due to forces associated with anisotropic radiation 
apparently results in negligible coagulation in the diffuse ISM.  

Gas-grain drift can also lead to variations in the dust-to-gas ratio.
If the size of a cold neutral medium region is $\sim 1 {\rm pc}$, then
the time to separate micron-sized grains from the region could be as short
as $\approx 2 \times 10^7 \yr$, which is perhaps comparable to the cloud
lifetime.  

\section{Summary}

We have investigated two forces that act on dust grains exposed to 
anisotropic radiation fields, due to the asymmetric photon-stimulated
emission of particles from the grains.  First, we have improved on 
Lafon's (1990) treatment of the photoelectric force, due to the recoil 
associated with photoelectrons, by using the realistic prescription of 
Kerker \& Wang (1982) for estimating the emission asymmetry and employing
the recent photoemission model of Weingartner \& Draine (2001b).

\begin{figure}
\epsscale{1.00}
\plotone{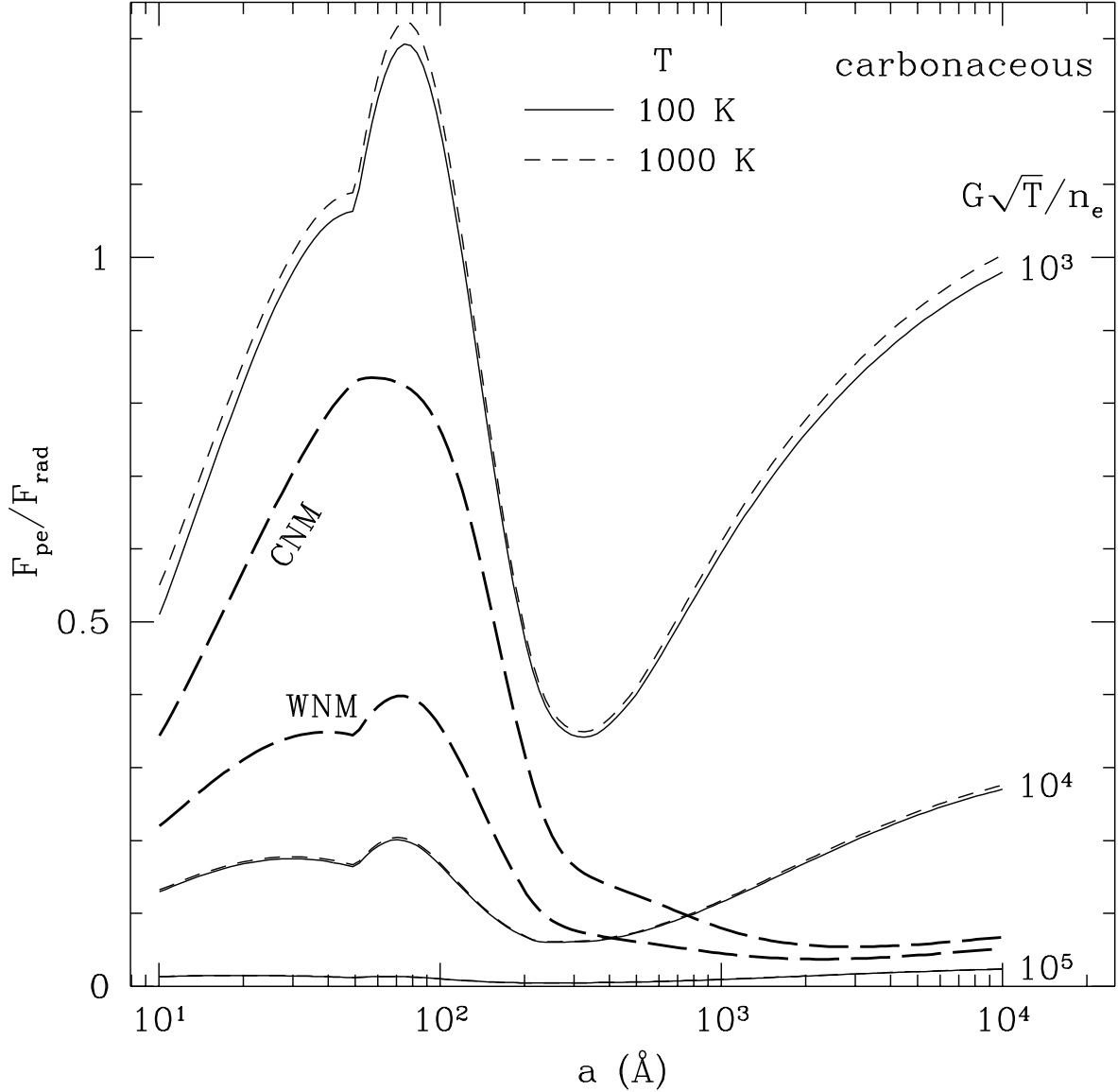}
\caption{
\label{fig:fpegra}
	$\Fpe/\Frad$ for carbonaceous grains, $\Tc = 3 \times 10^4 \K$,
and two gas temperatures:
$100 \K$ (solid) and $1000 \K$ (dashed).  The values of 
$G \sqrt{T} / n_e$, in ${\K}^{1/2}{\cm}^3$, are indicated.
Curves labelled CNM (WNM) are for cold (warm) neutral media and the 
ISRF.
        }
\end{figure}

\begin{figure}
\epsscale{1.00}
\plotone{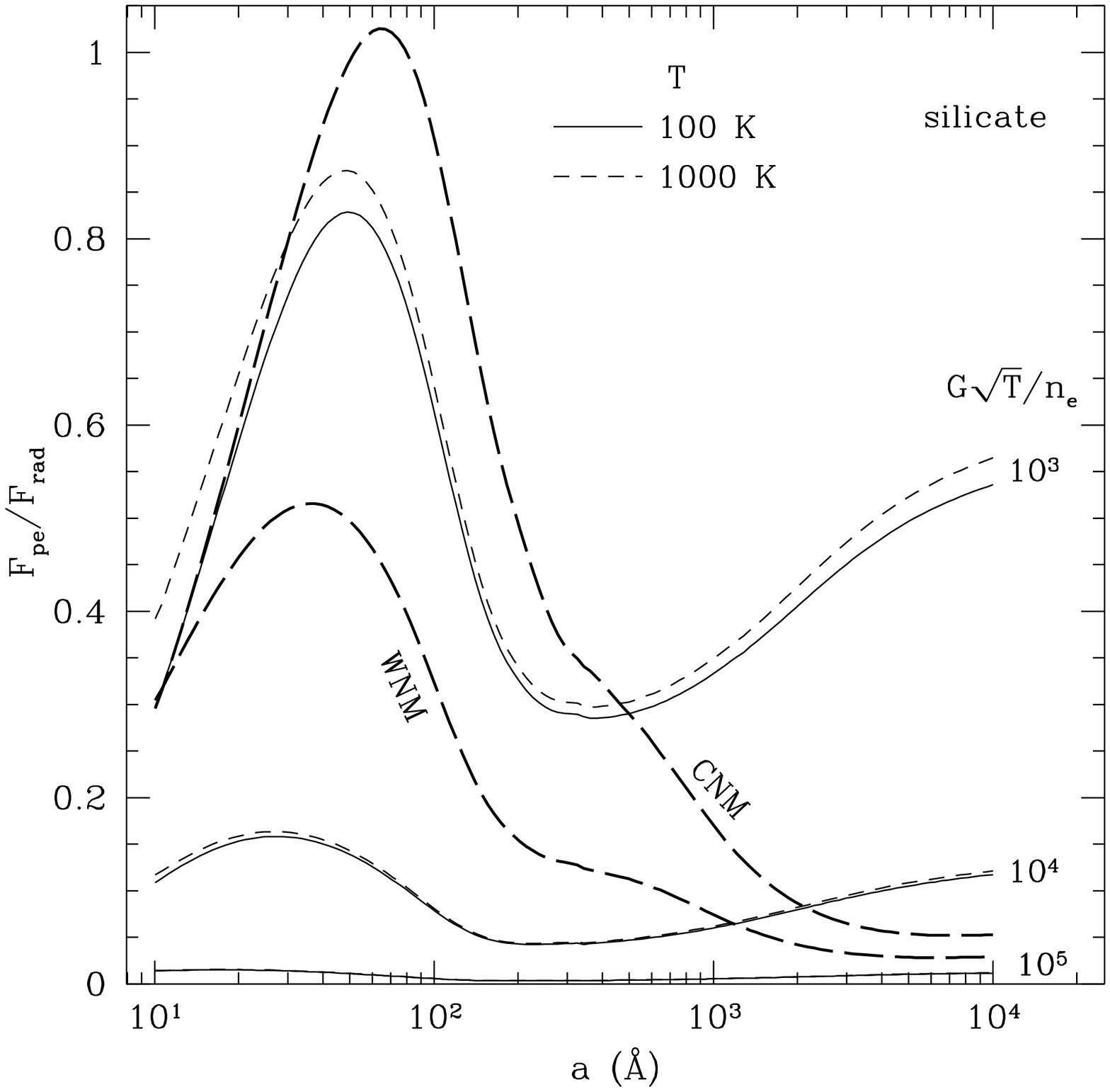}
\caption{
\label{fig:fpesil}
  	Same as Figure \ref{fig:fpegra}, but for silicate grains.
        }
\end{figure}

\begin{figure}
\epsscale{1.00}
\plotone{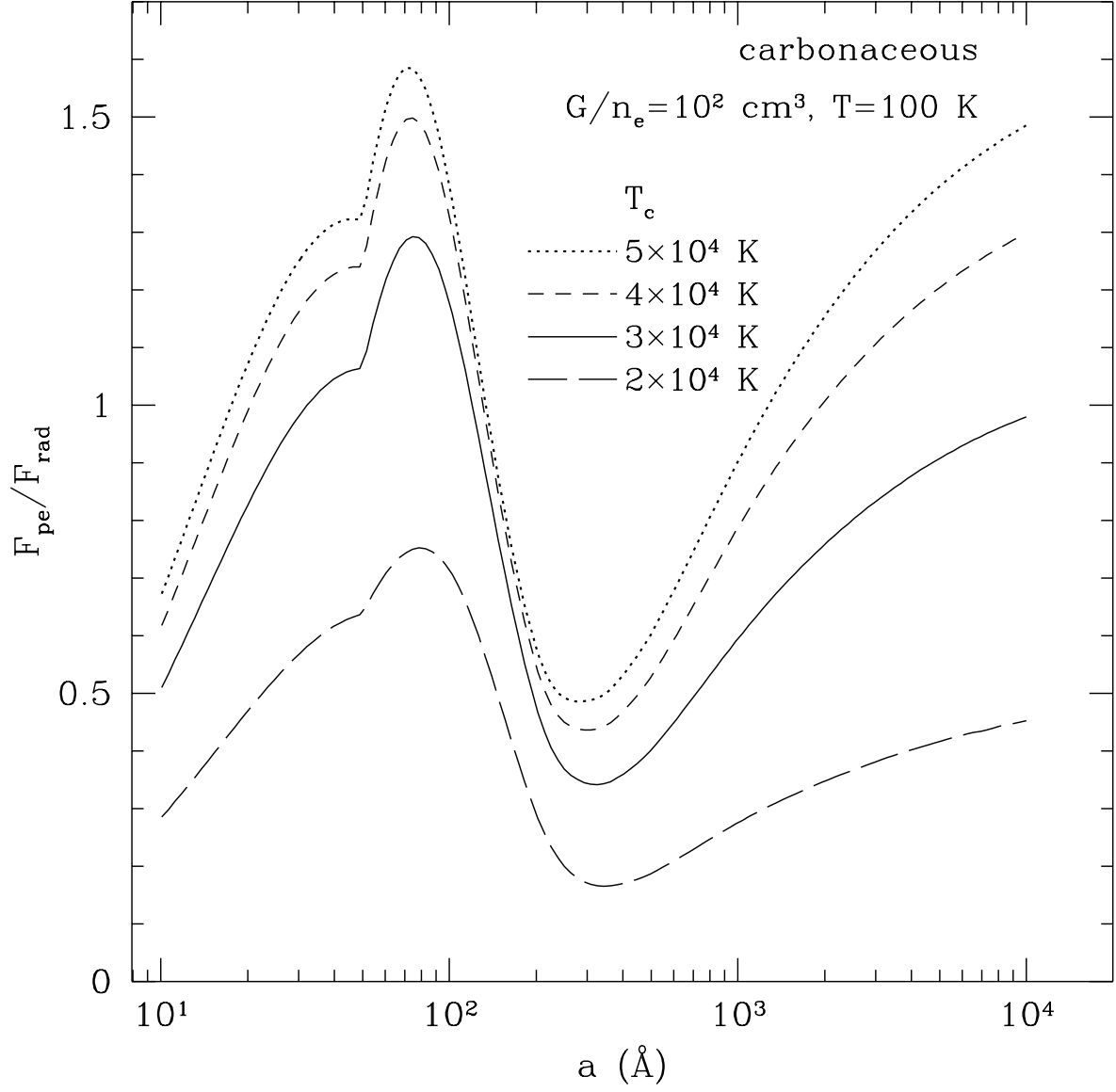}
\caption{
\label{fig:fpetc}
	$\Fpe/\Frad$ for carbonaceous grains, $T = 100 \K$, 
$G \sqrt{T} / n_e = {10}^3 {\K}^{1/2}{\cm}^3$, and four values of
$\Tc$ as indicated.  In all cases, the radiation field is assumed to be 
cut off at $13.6 \eV$.
        }
\end{figure}

The photoelectric force depends on the ambient conditions relevant
to the grain charging, including the shape of the radiation spectrum and the
parameter $G \sqrt{T}/n_e$.  We find that the photoelectric force is 
comparable to the radiation pressure in regions where the grain potential is
relatively low and the radiation spectrum is relatively hard (Figures 
\ref{fig:fpegra}--\ref{fig:fpetc}).  We find that the photoelectric force is
unimportant compared with the radiation pressure in the cold and warm 
neutral media for grains larger than a few hundred $\Angstrom$, because of 
the softness of the interstellar radiation field, as estimated by 
Mezger et al. (1982) and Mathis et al. (1983).  

We have also investigated the force resulting from the asymmetric 
photodesorption of H atoms adsorbed on the grain surface.  The surface 
physics and chemistry of likely grain materials are not currently 
well-known.  We have adopted a very simple, yet plausible, model in order to
examine the potential importance of this force.  In particular, we assume
that H chemisorbs at all surface sites with sufficient binding energy to 
supress thermal desorption and site-to-site diffusion.  

In our model,
the total force associated with gas-phase atoms which collide with the 
surface includes contributions from three processes:  photodesorption, 
H$_2$ formation, and elastic reflection of atoms incident on the grain
surface.  We call the force on a grain that is not drifting with respect
to the gas the ``photodesorption'' force.  When the grain drifts,
the total force (due to the above three processes) minus the photodesorption 
force equals the drag force, which is thus modified from the traditional
treatment considering only the reflection of incident atoms (\S 
\ref{sec:drag}).   

\begin{figure}
\epsscale{1.00}
\plotone{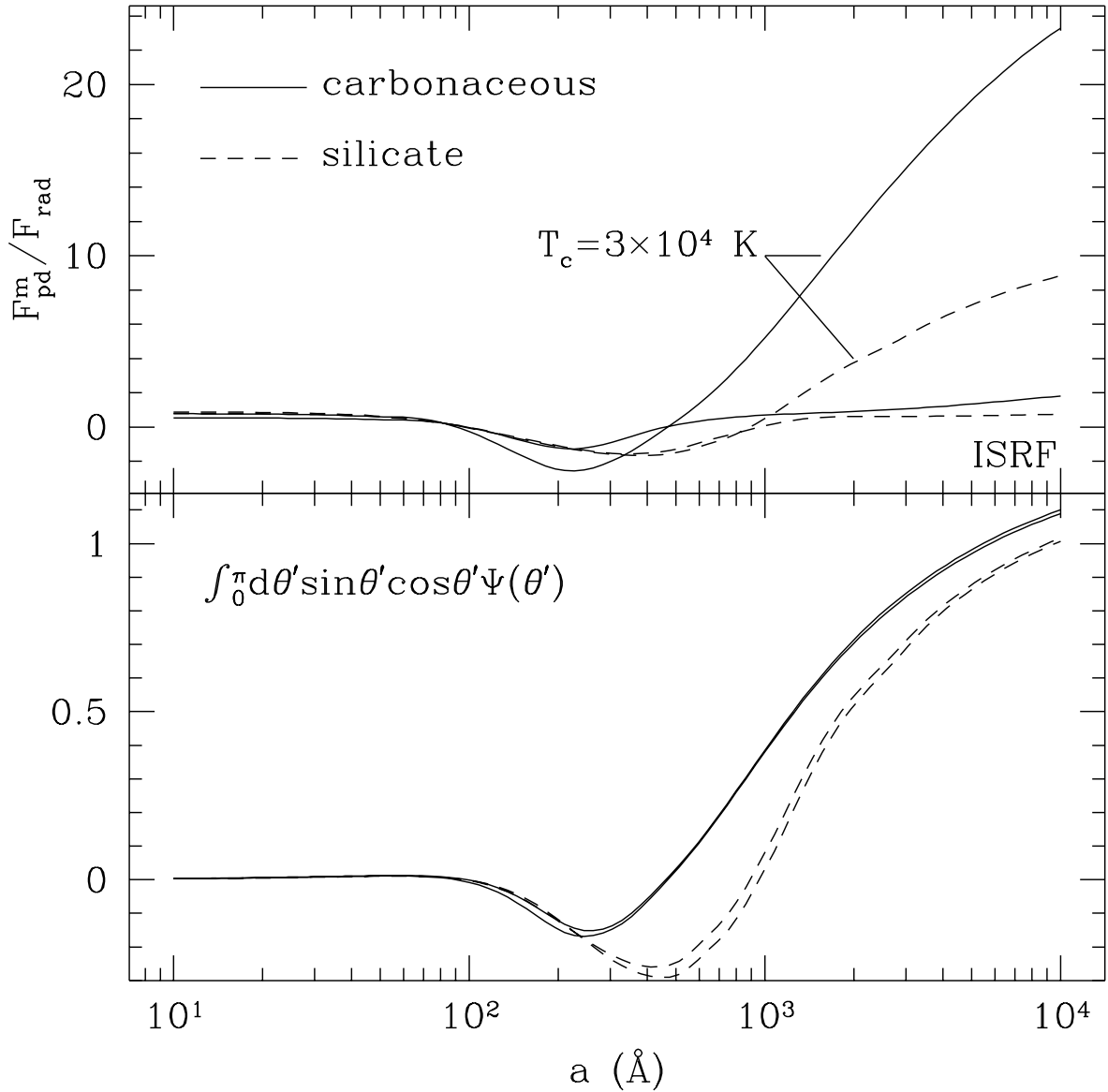}
\caption{
\label{fig:fpd1}
Lower panel:  The photodesorption asymmetry integral appearing in the 
expression for the maximum possible force for flipping grains, 
$F_{\rm pd}^{\rm m}$ 
(eq. \ref{eq:fpdm}).  For carbonaceous (silicate) grains, 
the curve corresponding to a blackbody radiation field 
with $T_c=3\times 10^4 \K$ (cut off at $13.6 \eV$) lies below (above) that for
the ISRF.  Upper panel:  Ratio of the maximum possible photodesorption force 
to the radiation pressure force, for parameter choices as indicated in 
Table \ref{tab:pdpars} (note that $E_{{\rm H}_2}$ is not relevant here).
        }
\end{figure}

With our standard choices for parameter values (Table \ref{tab:pdpars}),
we find that the photodesorption force dominates the radiation pressure
for grains with $a \gtsim 0.1 \micron$ when the radiation field is relatively
hard (Figure \ref{fig:fpd1}).  Laboratory studies of the surfaces of 
candidate grain materials, including determinations of photodesorption cross
sections, are needed in order to clarify the true importance of this force.

Grains exposed to anisotropic radiation will drift relative to the gas due to 
the combined action of the radiation pressure, photoelectric force, and 
photodesorption force.  We find that the radiation field in the
diffuse ISM is $\sim 10\%$ anisotropic in the visible and ultraviolet
(\S \ref{sec:anisotropy}).  In \S \ref{sec:drift_diffuse}, we estimate 
the maximum likely drift speeds for
grains in the cold and warm neutral media.  We find the timescale
for grain coagulation resulting from this drift to be $\gtsim 5 \times
10^9 \,$yrs.  Large grains ($a \sim 1 \micron$) might drift rapidly enough
to be substantially separated from the gas in cold diffuse clouds.  In a 
separate paper (Weingartner \& Draine 2001c), we will investigate 
gas-grain drift in photodissociation regions.

\acknowledgements
This research was supported in part by NSF grant
AST-9619429 and by NSF Graduate Research and International 
Fellowships to JCW.
We are grateful to R. H. Lupton for the availability of the SM plotting 
package.


\clearpage


\begin{deluxetable}{cc}
\tablecaption{Momentum Transfer Rates, Per Site \label{tab:momtransfer}}
\tablehead{
\colhead{process}&
\colhead{rate}
}
\startdata
photodesorption &$\frac{2}{3} R_{\rm pd} \pH f$ \\
H$_2$ formation &$\frac{2}{3} \Rao \cc \pHt f$ \\
reflections &$\left[1-\sH +f (\sH - \cc) \right] \int_0^{\infty} dv
P(v) \nH l^2 v (2 m_p v) = \left[1-\sH +f (\sH - \cc) \right]
m_p \Rao \sqrt{\frac{\pi}{\beta}}$ \\
sticking &$\left[ \sH - f(\sH-\cc) \right] \int_0^{\infty} dv P(v) \nH
l^2 v (m_p v) = \left[ \sH -f(\sH-\cc) \right] \frac{m_p \Rao}{2}
\sqrt{\frac{\pi}{\beta}}$ \\
\enddata
\end{deluxetable}

\begin{deluxetable}{ccc}
\tablecaption{Canonical Photodesorption Parameters \label{tab:pdpars}}
\tablehead{
\colhead{parameter}&
\colhead{symbol}&
\colhead{value}
}
\startdata
photodesorption rate (graphite) &
$\Ro$ &$5 \times 10^{-10} G \, {\rm s}^{-1}$ \\
photodesorption rate (silicate) &
$\Ro$ &$2 \times 10^{-10} G \, {\rm s}^{-1}$ \\
kinetic energy of photodesorbed H & $E_{\rm H}$ &$2 \eV$ \\
kinetic energy of newly-formed H$_2$ &$E_{{\rm H}_2}$ &$1 \eV$ \\
surface site area &$l^2$ &$6 \Angstrom^2$ \\
sticking probability &$\sH$ &1 \\
combination probability &$\cc$ &1 \\
\enddata
\end{deluxetable}

\begin{deluxetable}{ccccc}
\tablecaption{UV Energy Densities and Dipole Moments
\label{tab:uv}}
\tablehead{
\colhead{Quantity}&
\colhead{$2740 \Angstrom$}&
\colhead{$2365 \Angstrom$}&
\colhead{$1965 \Angstrom$}&
\colhead{$1565 \Angstrom$}
}
\startdata
starlight dipole ($p_s \nu u_{\nu}$)$_{\ast}$ \tablenotemark{a,b}
&$1.27$ &$0.843$ &$1.30$ &$1.67$ \\
total starlight ($\nu u_{\nu}$)$_{\ast}$ \tablenotemark{a,b}
&$4.43$ &$3.23$ &$4.73$ &$5.27$ \\
total $\nu u_{\nu}$ \tablenotemark{a,c}
&$4.90$ &$5.47$ &$6.47$ &$6.90$ \\
ISRF $\nu u_{\nu}$ \tablenotemark{a,d}
&$8.30$ &$6.20$ &$7.00$ &$8.20$ \\
dipole/ISRF &$0.152$ &$0.136$ &$0.185$ &$0.205$ \\
\enddata
\tablenotetext{a}{$10^{-14} \erg \cm^{-3}$}
\tablenotetext{b}{Gondhalekar 1989}
\tablenotetext{c}{Gondhalekar et al. 1980}
\tablenotetext{d}{Mathis et al. 1983}
\end{deluxetable}

\begin{deluxetable}{ccccc}
\tablecaption{Anisotropy Directions
\label{tab:dir}}
\tablehead{
\colhead{Band}&
\colhead{RA}&
\colhead{dec}&
\colhead{$l$}&
\colhead{$b$}
}
\startdata
$1565 \Angstrom$ &$141^{\circ}$ &$-62^{\circ}$ &$281^{\circ}$ 
&$-8.2^{\circ}$ \\
$1965 \Angstrom$ &$132^{\circ}$ &$-58^{\circ}$ &$275^{\circ}$
&$-8.9^{\circ}$ \\
$2365 \Angstrom$ &$124^{\circ}$ &$-55^{\circ}$ &$270^{\circ}$
&$-10.5^{\circ}$ \\
$2740 \Angstrom$ &$129^{\circ}$ &$-56^{\circ}$ &$273^{\circ}$
 &$-9.3^{\circ}$ \\
$B$ &$115^{\circ}$ &$-43^{\circ}$ &$257^{\circ}$ &$-9.8^{\circ}$ \\
$V$ &$115^{\circ}$ &$-51^{\circ}$ &$264^{\circ}$ &$-13.6^{\circ}$ \\
\enddata
\tablecomments{
        Celestial coordinates:  right ascension (RA) and declination (dec);
Galactic coordinates:  longitude ($l$) and latitude ($b$).
        }
\end{deluxetable}

\end{document}